\documentclass[
  journal=largetwo,
  manuscript=research-paper,
  year=2024,
]{cup-journal}

\usepackage{amsmath}
\usepackage{geometry}
\usepackage{graphicx,graphics}	% Including figure files
\usepackage[T1]{fontenc}
\usepackage{pbox}
\usepackage{booktabs}
\usepackage{tabularx}
\usepackage{setspace,hyperref}
\usepackage[utf8]{inputenc}
\usepackage{subcaption}

\DeclareUnicodeCharacter{0301}{\hspace{-1ex}\'{ }}

\usepackage{hyperref}

\title{Clustering and physical properties of AGN and Star-Forming Galaxies at fixed stellar mass: does assembly bias have a role in AGN activity?}

\author{Amrita Banerjee}
\affiliation{Centre for Astrophysics and Supercomputing, Swinburne University of Technology, PO Box 218, Hawthorn, VIC 3122, Australia}

\author{Biswajit Pandey}
\affiliation{Department of Physics, Visva-Bharati University, Santiniketan, 731235, West Bengal, India}
\email[Biswajit Pandey]{biswap@visva-bharati.ac.in}
\author{Anindita Nandi}
\affiliation{Department of Physics, Visva-Bharati University, Santiniketan, 731235, West Bengal, India}

\keywords{methods: data analysis, methods: statistical, galaxies: statistics}

\begin{document}

\begin{abstract}
  We analyze a volume-limited sample from the Sloan Digital Sky Survey
  (SDSS) to compare the spatial clustering and physical properties of
  active galactic nuclei (AGN) and star-forming galaxies (SFG) at
  fixed stellar mass. We find no statistically significant difference
  in clustering strength or local density between AGN and
  SFG. However, after matching their stellar mass distributions, we
  detect statistically significant differences (at a confidence level
  $>99.99\%$) in colour, star formation rate (SFR),
  $4000\textup{~\AA}$ break measurements (D$4000$), and
  morphology. These differences persist across both low- and
  high-density environments, suggesting that AGN are not driven by
  environmental factors. The development of favourable conditions for
  AGN activity within a galaxy may depend on the diverse evolutionary
  histories of galaxies. Our results imply that AGN activity may arise
  stochastically, modulated by the complex assembly history of
  galaxies.
\end{abstract}

%**************************************************************************************************************** %
\section{Introduction}
\label{sec:intro}

AGN rank among the brightest astrophysical
sources in the universe, emitting radiation across the entire
electromagnetic spectrum with bolometric luminosities around
$10^{47}-10^{48}$ erg/s \citep{fabian99, woo02}. This intense
radiation is believed to stem from the accretion of matter onto
supermassive black holes (SMBH) located at the centers of massive
galaxies. As gas clouds spiral toward the SMBH, losing angular
momentum, their gravitational potential energy is converted into
electromagnetic radiation \citep{jiang13, cielo18}. This radiation can
then heat the surrounding gas, hindering its cooling and delaying star
formation \citep{kawata05, antonuccio10, wagner13}. Additionally,
energy and momentum from AGN-driven outflows and radio jets can either
heat or expel gas \citep{morganti17, baron18, santoro20}, thereby
limiting black hole growth and suppressing further star formation.

AGN feedback is widely regarded as fundamental to the co-evolution of
galaxies and their central black holes \citep{somerville08,
  kormendy13, heckman14, harrison17}. Observations indicate a decline
in the star formation rate after $z \sim 1$ \citep{madau96, hopkins04,
  behroozi13}. The observed bimodality in the colour distribution
\citep{strateva01, blanton03, balogh04, baldry04, pandey20} indicates
that the galaxies are transitioning from the actively star-forming
blue population to a passively evolving red sequence. The exact
physical processes driving this transition, particularly the quenching
of star formation in the transitional ``green valley'', remain
uncertain \citep{das21}. However, numerous studies propose that AGN
feedback may play a crucial role in quenching star formation in this
phase \citep{nandra07, hasinger08, silverman08, cimatti13,
  zhang21}. The models of galaxy formation and evolution increasingly
rely on AGN feedback to replicate observed galaxy properties, making
it an essential element in theoretical, numerical, and semi-analytic
models \citep{springel05, dimatteo05, eckert21}.

Nearly all massive galaxies harbour a supermassive black hole (SMBH)
at their center, yet only a subset exhibit AGN activity at any given
time. Understanding what triggers AGN activity in these galaxies is
critical, as various internal and external factors shape the
likelihood of such activity. Internal characteristics, such as gas
availability in the central region, host galaxy kinematics, and
morphology, significantly influence the accretion of gas onto the
central SMBH \citep{ruffa19, shangguan20, ellison21,
  sampaio23}. Additionally, the mass of the host dark matter halo
affects gas reservoir availability and the galaxy's capacity to draw
gas from its surroundings. Larger halos, with deeper potential wells,
facilitate gas inflow toward the galactic center, thus making AGN
activity more probable \citep{georga19, aird21, luo22}. Observational
data also reveal that AGN activity is more frequent in massive
galaxies \citep{dunlop03, brusa09, pimbblet13}.

The SMBH mass itself plays a vital role in AGN dynamics. The larger
black holes exert stronger gravitational forces, enabling higher
accretion rates and boosting AGN luminosity. Meanwhile, AGN feedback
can limit black hole growth by modulating gas supply. Massive
galaxies, often found in high-mass dark matter halos within dense
environments like galaxy clusters and cosmic web filaments, may
experience indirect influence from these environments. Observations
suggest that galaxy colour and star formation rates are sensitive to
cosmic web environments \citep{pandeysarkar20, das23a,
  das23b}. Furthermore, gas inflow along cosmic web filaments can
initiate and sustain AGN activity within galaxies \citep{umehata19}.

Numerous studies indicate that AGN are more strongly clustered than
SFG \citep{gilli09, mandelbaum09, kollat12, donoso14, hale18}. Using
SDSS data, \citet{satyapal14} observe that the fraction of AGN
increases as the distance to neighbouring galaxies
decreases. Similarly, \citet{zhang21} find that AGN have more
neighbouring galaxies compared to SFG. Results from the Horizon Run 5
simulation \citep{lee21}, as analyzed by \citet{singh23}, show that
AGN activity rises in response to both higher background densities and
closer proximity to neighbouring galaxies. Physical mechanisms,
including major and minor mergers \citep{dimatteo05, alonso07,
  ellison11, storchi19}, disk instability \citep{hopkins06, dekel09,
  hopkins14}, and tidal effects \citep{moore96}, are thought to
enhance the supply of cold gas to the central SMBH, thereby boosting
AGN activity. Interactions and mergers, more common in clusters and
filaments, often drive gas inflows toward galactic centers, further
promoting AGN activity \citep{hernquist89, springel05, alexander12}.

While AGN are generally more common in dense environments, extremely
high-density regions like massive galaxy clusters present a more
complex picture. The pressure from the hot intracluster medium (ICM)
at the centers of massive galaxy clusters can cause ram pressure
stripping of the cold gas that fuels the AGN activity \citep{gunn72,
  abadi99, boselli22}. Additionally, the cluster halo may capture the
cold gas, preventing accretion towards the inner regions by
strangulation \citep{larson80, peng15}. These processes often suppress
AGN activity near the centers of massive galaxy
clusters. \citet{ehlert14} observe that the fraction of X-ray bright
AGN rises with increasing distance from the centers of galaxy
clusters, and \citet{lopes17} find that AGN are more frequently located
in low-mass groups, field environments, and cluster outskirts. The XXL
survey \citep{pierre16}, as analyzed by \citet{koulouridis18}, reveals
that the relationship between X-ray-selected AGN and environment
differs between high- and low-mass clusters. Studies of X-ray selected
clusters from ROSAT by \citet{mishra20} show a lower AGN fraction in
clusters compared to fields, while \citet{cecca21} find AGN activity
significantly stronger in voids compared to field environments.

Low-density regions, such as voids, tend to host less evolved galaxies
due to the lack of external processes, like gas stripping and frequent
mergers, and contain large reservoirs of pristine gas. Galaxies in
these environments evolve through internal, or secular, processes and
are typically fainter, bluer, and exhibit higher star formation rates
than galaxies in average-density environments \citep{grogin00,
  hoyle05, ricciardelli14, bruton20}. \citet{constantin08} find that
moderately luminous AGN are more common in voids than walls, but the
abundance of brighter AGN are comparable in the two
environments. \citet{kauffmann03} observe a decreasing AGN fraction in
massive galaxies as density increases, and several other works report
a higher prevalence of AGN in low- to moderate-density environments
\citep{kauffmann04, gilmour07, choi09, sabater13, miraghaei20,
  mishra21}. This trend suggests that galaxies in voids may experience
a higher frequency of one-on-one interactions, which may be key to
triggering AGN activity in these regions.

The environmental dependence of AGN activity at higher redshifts has
been investigated in several studies. Using data from the zCOSMOS
spectroscopic survey up to $z \sim 1$, \citet{silverman09} find that
massive galaxies hosting AGN tend to reside in low-density regions. In
contrast, \citet{bradshaw11} analyze the UKIDSS Ultra-deep Survey in
the redshift range $z \sim 1-1.5$ and observe that AGN are more
frequently found in high-density environments. More recent studies
provide growing evidence for a positive evolution of AGN activity with
redshift, particularly in dense environments such as galaxy
clusters. Several works have demonstrated that the fraction of AGN in
clusters increases with redshift, implying a stronger connection
between environment and AGN triggering at earlier times
\citep{fassbender12, martini13, bufanda17, hasiguchi23}. Numerous
studies also reported a high incidence of AGN in proto-cluster
environments, further supporting the idea that dense regions at high
redshift are conducive to AGN activity \citep{lehmar13, krishnan17,
  gatica24, vito24}.

Several other studies suggest that AGN activity shows little to no
dependence on environmental factors. \citet{miller03} report that the
fraction of optically selected AGN remains consistent from the cores
of galaxy clusters to field regions, a finding mirrored by
\citet{martini07} for X-ray-selected AGN. Similarly, \citet{pandey08}
analyze SDSS data, comparing filamentarity in the distributions of
SFG and AGN, and find no significant
difference. \citet{pimbblet13} observe that the fraction of optically
selected AGN does not vary with distance from the cluster center,
while \citet{sabater15} find no statistically significant effect of
environment on optical AGN activity. Likewise, \citet{amiri19} report
only a weak correlation between local galaxy density and AGN activity,
and \citet{man19}, analyzing SDSS data, find minimal to no
environmental influence on AGN occurrence. Some studies find no
significant differences in the clustering of AGN and non-AGN galaxies
\citep{porqueres18, wang19}. These apparently conflicting results
suggest that the role of environment in AGN activity remains an open
question, underscoring the need for further research to resolve the
uncertainties.

The SDSS \citep{stout02} provides high-quality spectra and imaging for
a large number galaxies in the nearby universe, making it one of the
largest and most comprehensive redshift surveys to date. The precise
classification of SFG and AGN based on emission lines enables a robust
statistical comparison between these populations. In this study, we
investigate the clustering properties of SFG and AGN using statistical
tools such as the two-point correlation function and nearest neighbour
distribution. The mass of a galaxy is known to influence the AGN
activity. However, galaxy mass is known to depend on environment. To
identify any additional dependence of AGN activity on the environment,
we will compare the spatial clustering and physical properties of the
SFG and AGN after matching their stellar mass distributions. This
approach will allow us to assess any residual environmental impact on
AGN activity by comparing the spatial clustering and physical
properties of SFG and AGN at similar masses across varying
densities. Additionally, analyzing AGN and SFG properties in different
environments at the same mass could clarify the influence of
large-scale environment and assembly bias \citep{gao05, wechsler06,
  gao07, croton07}. The dark matter halos of similar mass may have
distinct assembly histories leading to different halo concentration,
merger rates, or gas accretion rates, potentially impacting the AGN
activity. In this study, we will explore the possible roles of
assembly bias on the AGN activity in galaxies.

We use a $\Lambda$CDM cosmological model with $\Omega_{m0}=0.315$,
$\Omega_{\Lambda0}=0.685$ and $h=0.674$ \citep{planck20} throughout
the present work.

The outline of our work is as follows. In Section 2, we describe our
data and the methods of analysis. Section 3 presents our results, and
in Section 4, we provide our conclusions.

% ***************************************************************************************************************** %
\section{Data and method of analysis}
\label{sec:datamethod}

\subsection{SDSS data}
We use data from the $17^{th}$ data release (DR17) of SDSS
\citep{abdur22}, which is a multi-band imaging and spectroscopic
redshift survey. The SDSS employs a $2.5\,\rm m$ optical telescope
\citep{gunn06} at Apache Point Observatory in New Mexico, USA, to
gather photometric and spectroscopic data on galaxies across
one-quarter of the entire sky. DR17 covers $14555~$ square degrees and
includes spectroscopic information for $2863635$ galaxies. For our
analysis, we focus on the Main Galaxy Sample \citep{strauss02} of the
SDSS. The data are accessed via the SDSS CasJobs
service\footnote{\url{https://skyserver.sdss.org/casjobs/}} using
Structured Query Language (SQL).

We select a contiguous region of the sky in equatorial coordinates,
specifically the area spanning $130^\circ \leq \alpha \leq 230^\circ$
and $0^\circ \leq \delta \leq 60^\circ$, for our analysis. From this
region, we download the spectroscopic data for galaxies with redshifts
in the range $0 \leq z \leq 0.2$ and r-band apparent Petrosian
magnitudes $m_r < 17.77$. These criteria yield a total of $392292$
galaxies.

We use the \textit{galSpecExtra} table, which is derived from the
MPA-JHU spectroscopic catalog of
galaxies \footnote{\url{https://www.sdss4.org/dr17/spectro/galaxy_mpajhu/}},
to classify the objects as AGN or SFG based on the BPT diagram
\citep{brinchmann04}. In this table, SFG are flagged with a value of
1, while AGN are flagged with a value of 4. Our AGN sample primarily
consists of high signal-to-noise ratio (SNR) narrow line AGN excluding
the composite galaxies (flag 3) and galaxies with low-ionization
nuclear emission-line regions (LINERs). The information about the
stellar mass and SFR are also provided in the \textit{galSpecExtra}
table. The stellar masses of the galaxies are estimated using the
methodology outlined in \cite{kauffmann03}, applied to photometric
data as detailed in \cite{salim07}.  Star formation rates are
calculated according to the approach discussed in
\cite{brinchmann04}. The aperture corrections are made by estimating
star formation rates from SED fits to the photometry outside the
fiber, following the methodology described in
\cite{salim07}. Estmating SFR in AGN through model fitting gives
unreliable results since different lines are affected by AGN in
different ways. The sSFR for AGN in MPA-JHU catalogue are calculated
using D4000 values. The D$4000$, which indicate the mean age of the
stellar population in galaxies \citep{Balogh1999}, are retrieved from
the \textit{galSpecIndx} table. To characterize the morphology of
galaxies, we use the concentration index, $\frac{r_{90}}{r_{50}}$
\citep{shimasaku01}, where $r_{90}$ and $r_{50}$ represent the radii
that contain $90\%$ and $50\%$ of the Petrosian flux,
respectively. These values are obtained from the \textit{PhotoObjAll}
table.

\begin{figure}
    \centering
    \includegraphics[width = 7.5cm]{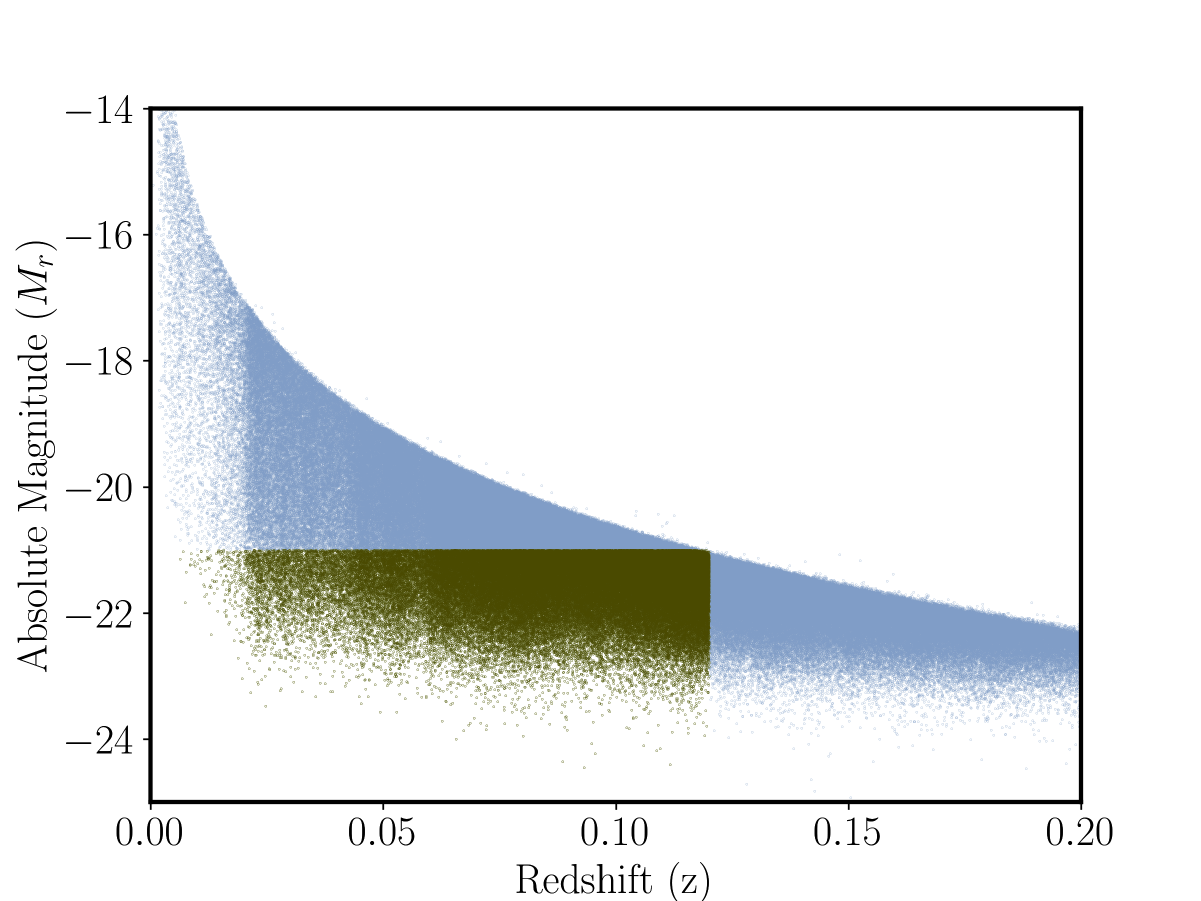}
    \caption{This shows the definition of the volume limited sample in
      the redshift-absolute magnitude plane. The volume limited sample
      comprises of the galaxies lying within the rectangular region in
      this diagram.}
    \label{fig:vls}
\end{figure}

We construct a volume-limited sample by applying a cut on the
K-corrected and extinction-corrected r-band absolute magnitude,
selecting galaxies with $M_r \leq -21$. This corresponds to a redshift
cut of $z \leq 0.12$. The resulting sample consists of a total of
$111671$ galaxies (\autoref{fig:vls}), which include $38606$
unclassified galaxies, $17282$ star-forming galaxies, $22943$ low
SNR star-forming galaxies, $10028$ composite
galaxies, $5828$ AGNs, and $16984$ low SNR LINERs.

\begin{figure*}
\centering
    \includegraphics[width = 14cm]{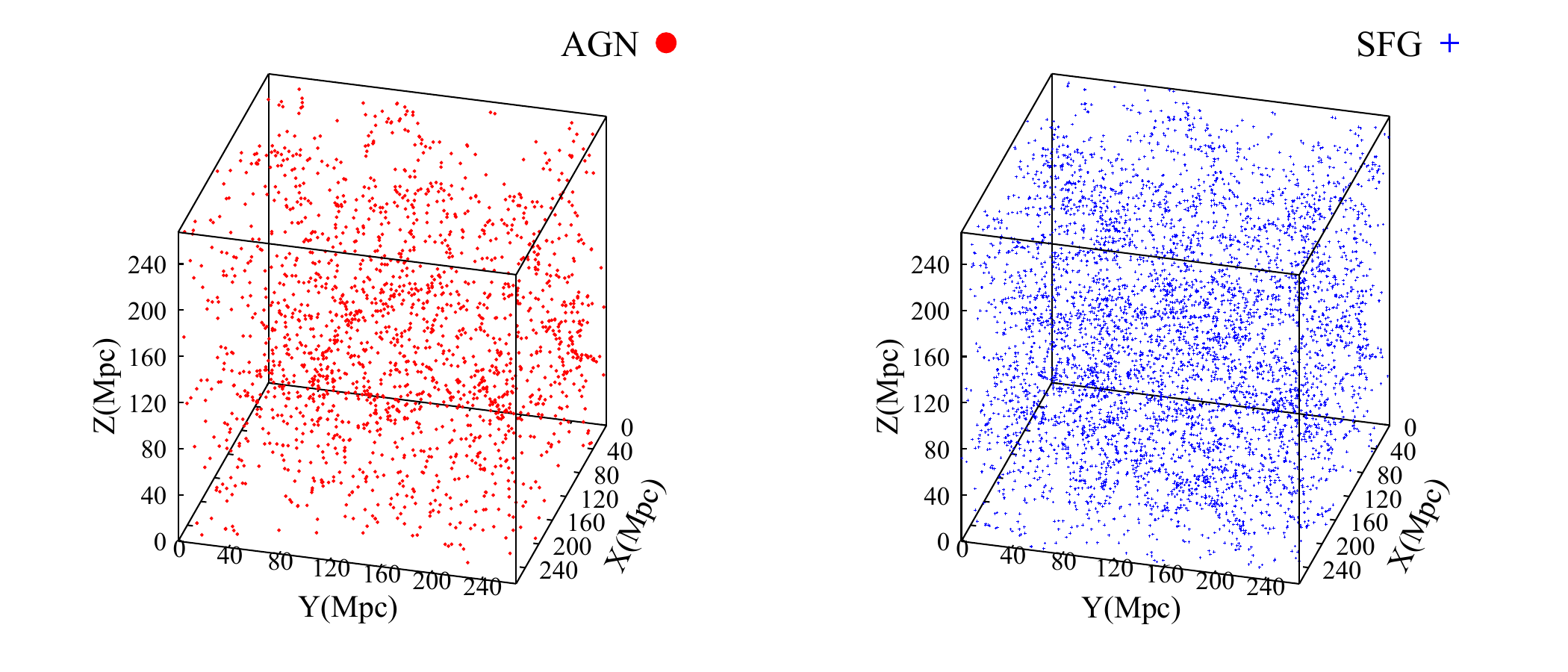}
    \caption{The left and right panels of this figure, respectively
      show the spatial distributions of the AGN and SFG within the
      datacube extracted from the volume limited sample.}
 \label{fig:dist}
\end{figure*}

We extract the largest cube that can be fitted within the
volume-limited sample. This datacube has a side length of $267.5$ Mpc
and contains a total of $30860$ galaxies, of which $5184$ are SFG and
$1883$ are AGN. The primary objective of this work is to compare the
spatial clustering and physical properties of AGN and SFG. Therefore,
we focus our analysis on these two galaxy types. The spatial
distributions of AGN and SFG within the extracted datacube are shown
in \autoref{fig:dist}.

%%%%%%%%%%%%%%%%%%%%%%%%%%%%%%%%%%%%%%%%%%%%%%%%%%%%%%%%%%%%%%%%%%%%%%%%%%%%%%%%%%

%%%%%%%%%%%%%%%%%%%%%%%%%%%%%%%%%%%%%%%%%%%%%%%%%%%%%%%%%%%%%%%%%%%%%%%%%%%%%%%%%%

\subsection{Methods of analysis}

\subsubsection{Matching the stellar mass distributions of the AGN and SFG}
\label{sec:matching}
\begin{figure*}[htbp!]
    \centering
    \includegraphics[width = 14cm]{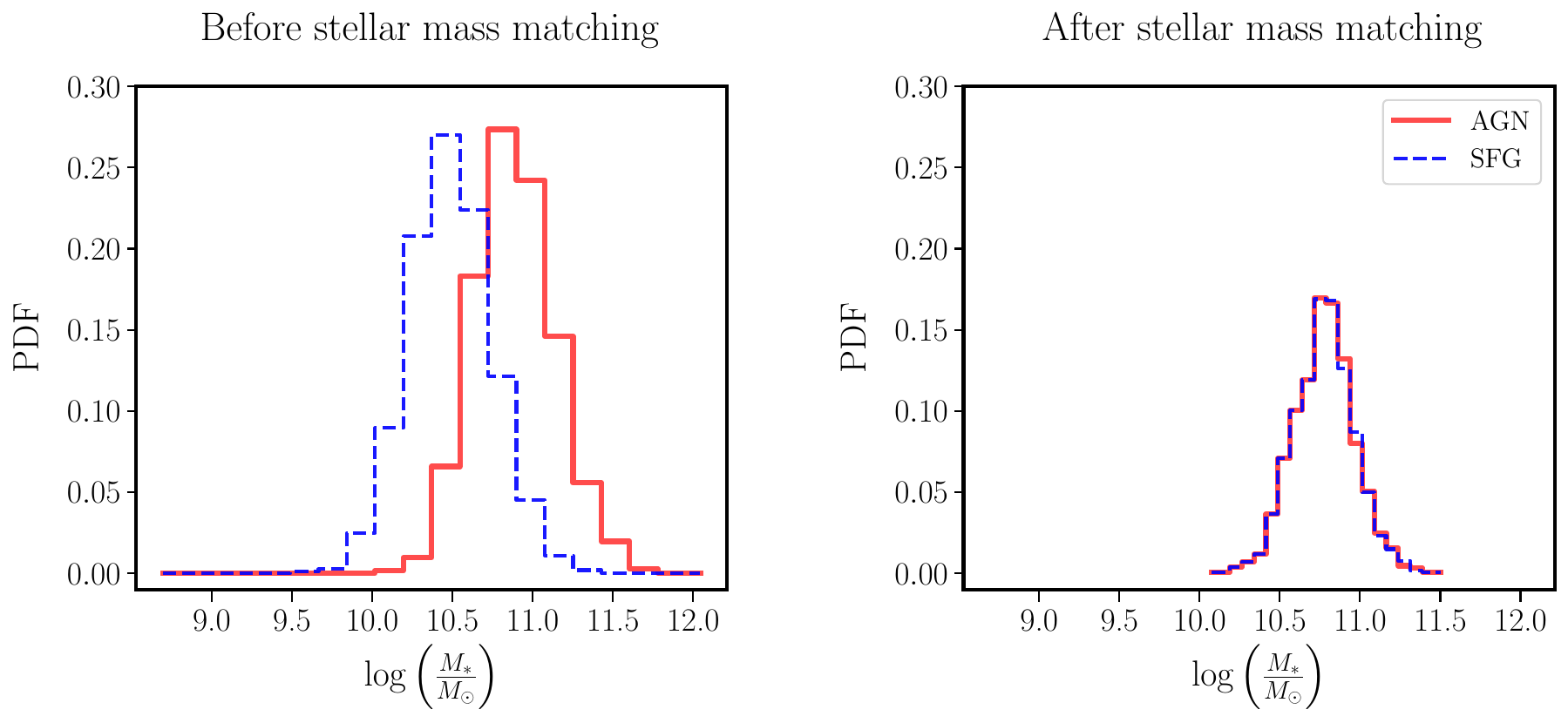}
    \caption{The left panel of this figure shows the stellar mass
      distributions of the AGN and SFG. We match the AGN and SFG
      stellar mass distributions, which are shown together in the
      right panel.}
 \label{fig:match}
\end{figure*}

The stellar mass of a galaxy is a key factor influencing the onset of
AGN activity. AGN abundance tends to
increase with the stellar mass of the host galaxy \citep{kauffmann03,
  silverman09}. This strong correlation between AGN activity and
galaxy mass could introduce significant bias into our study if not
properly accounted for. To address this, we match the stellar mass
distributions of AGN and SFG in our sample
using the criterion $|\frac{m_{SFG}}{m_{AGN}} - 1| < 10^{-3}$. The
stellar mass distributions for both AGN and SFG, before and after
matching, are shown in the left and right panels of
\autoref{fig:match}, respectively. We apply a Kolmogorov-Smirnov (KS)
test to compare the distributions after matching and find that the
null hypothesis can be rejected with very low confidence (p-value $<
1\%$).

We calculate the fraction of AGN as a function of stellar mass for the
galaxies in our datacube and present the results in
\autoref{fig:frac}. The left panel of \autoref{fig:frac} shows that
the AGN fraction $\frac{AGN}{AGN+SFG}$ increases steadily with stellar
mass for galaxies with masses greater than $\sim
10^{10.5}\,M_{\odot}$. The sharp rise in the ratio $\frac{AGN}{SFG}$
at higher masses, shown in the right panel of \autoref{fig:frac}, is
due to the lower abundance of SFG at these mass scales. Galaxies with
masses above $3 \times 10^{10}\,M_{\odot}$ are predominantly
quiescent, bulge-dominated galaxies, while those with lower masses are
typically actively star-forming and have disk-like morphologies
\citep{kauffmann03a}. Hydrodynamical simulations suggest that a
transition occurs around this critical mass from cold-mode to hot-mode
accretion, leading to mass quenching in galaxies \citep{birnboim03,
  dekel06, keres05, gabor10}. In these more massive galaxies, the halo
gas can eventually cool and collapse to form stars. However, AGN
feedback can provide additional heating, preventing this cooling and
maintaining a hot halo \citep{fabian12, mcnamara12}. The higher AGN
fraction observed in more massive galaxies suggests that these
galaxies provide a more conducive environment for AGN
activity. Moreover, the more massive galaxies are strongly clustered
and tend to reside in high-density regions. This implies that any
comparison of clustering between SFG and AGN would be influenced by
the mass dependence of clustering.

The primary goal of this study is to compare the clustering and
physical properties of SFG and AGN with similar stellar
masses. Since the environment, clustering, and physical properties of
galaxies are strongly influenced by its mass, we match the stellar
mass distributions of the two populations to ensure that our results
are not biased by mass-dependent factors. Although only about $20\%$
of the most massive galaxies in our SFG sample are available for
comparison with AGN, this approach allows us to explore the roles of
other potential factors, beyond stellar mass, that might contribute to
AGN activity.

%%%%%%%%%%%%%%%%%%%%%%%%%%%%%%%%%%%%%%%%%%%%%%%%%%%%%%%%%%%%%%%%%%%%%%%%%%%%%%%%%%
\subsubsection{Two-point correlation function}
\label{sec:corrfunc}
The two-point correlation function quantifies the strength of galaxy
clustering at a given scale by measuring the excess probability of
finding two galaxies at a specific separation, compared to a random
Poisson distribution. We calculate the two-point correlation function
separately for the AGN and SFG samples, after matching their
stellar mass distributions. The data extracted from the volume-limited
sample includes $1883$ AGN and $5184$ SFG. After stellar mass matching, we obtain $1285$ AGN and $1285$
SFG galaxies. 

The two-point correlation function is computed using the Landy and
Szalay estimator \citep{1993ApJLandy}:
\begin{equation}
    \xi(r) = \frac{DD(r) - 2DR(r)+ RR(r)}{RR(r)}
    \label{eq:2pcf_estimator}
\end{equation}
where $DD(r)$, $RR(r)$ and $DR(r)$ are normalized counts for
data-data, random-random and data-random pairs at separation $r$.
To estimate the error bars, we generate 50 jackknife
resamplings for each dataset.

%%%%%%%%%%%%%%%%%%%%%%%%%%%%%%%%%%%%%%%%%%%%%%%%%%%%%%%%%%%%%%%%%%%%%%%%%%%%%%%%%%
\subsubsection{Distribution of the $n^{th}$ nearest neighbour distance and the local density}
\label{sec:5thnn}
Galaxies in denser environments are expected to have closer
neighbours. The distance to the $n^{th}$ nearest neighbour, $r_{n}$,
can serve as a proxy for the local environment \citep{casertano85} of
a galaxy, with $n$ representing the number of neighbours
considered. In our analysis, we focus on three-dimensional space and
select $n=5$ for the present study.

We calculate the distribution of the $5^{th}$ nearest neighbour
distances for both AGN and SFG galaxies, using all $30860$ galaxies in
our dataset.

The local galaxy density around an AGN or SFG is defined as,
\begin{eqnarray}
{\eta}_n = \frac{n-1}{V(r_n)}  
\label{eqn:knn}
\end{eqnarray}  
where, $V(r_n)=\frac{4}{3}\pi r_{n}^3$ is the volume within a radius $r_n$.

Due to the sharp boundaries of our samples, the local density can be
underestimated near the edges. To address this, we calculate the
minimum distance $r_b$ from each galaxy to the boundary of the sample
and only include galaxies for which $r_n<r_b$ in our local density
calculations.

%%%%%%%%%%%%%%%%%%%%%%%%%%%%%%%%%%%%%%%%%%%%%%%%%%%%%%%%%%%%%%%%%%%%%%%%%%%%%%%%%%

\section{Results and Discussions}
\label{sec:result}

%\hfill
\begin{figure*}[htbp!]
    \centering
    \includegraphics[width = 7.5cm]{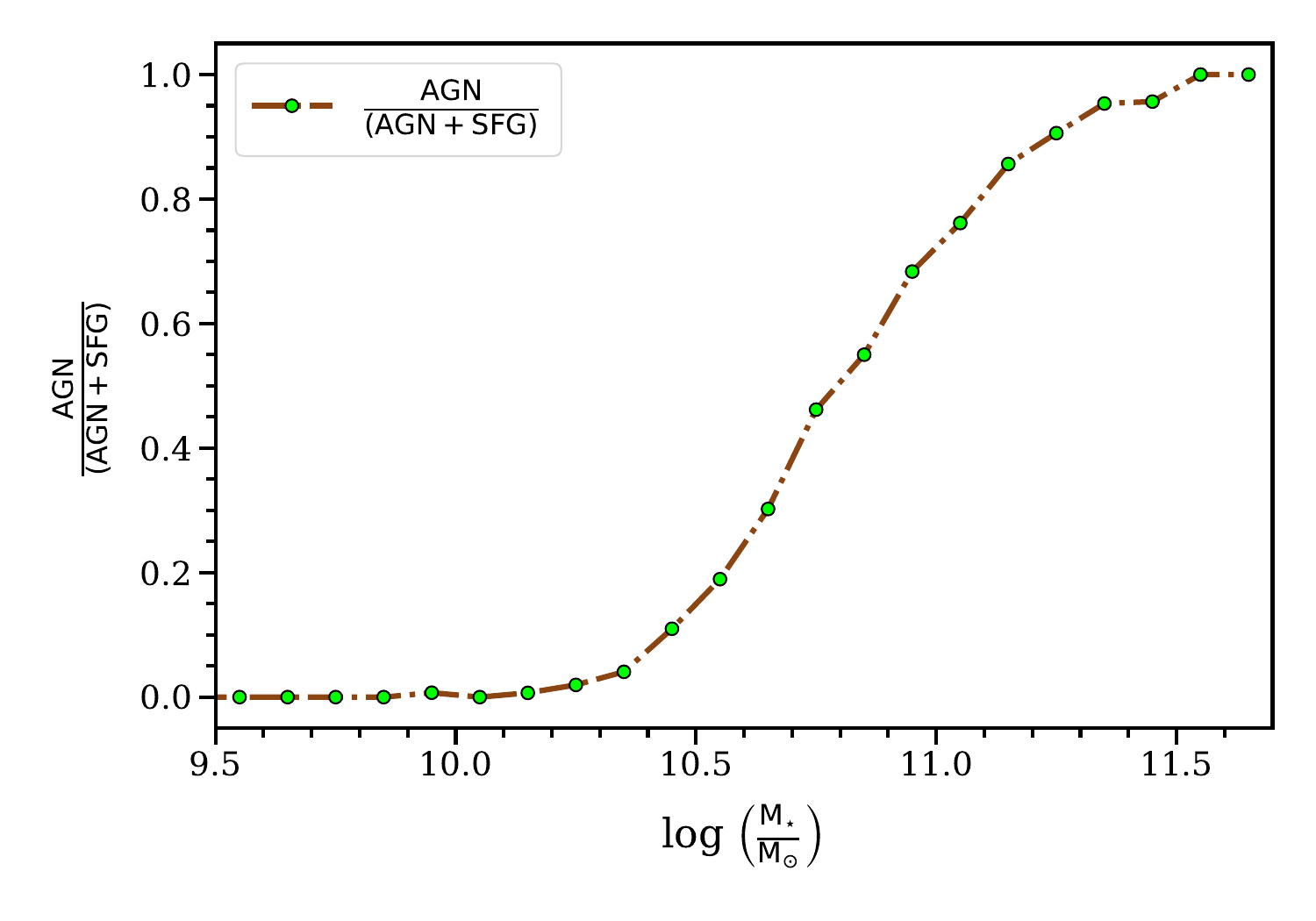}
%    \vspace{-0.4cm}
    \includegraphics[width=7.5 cm]{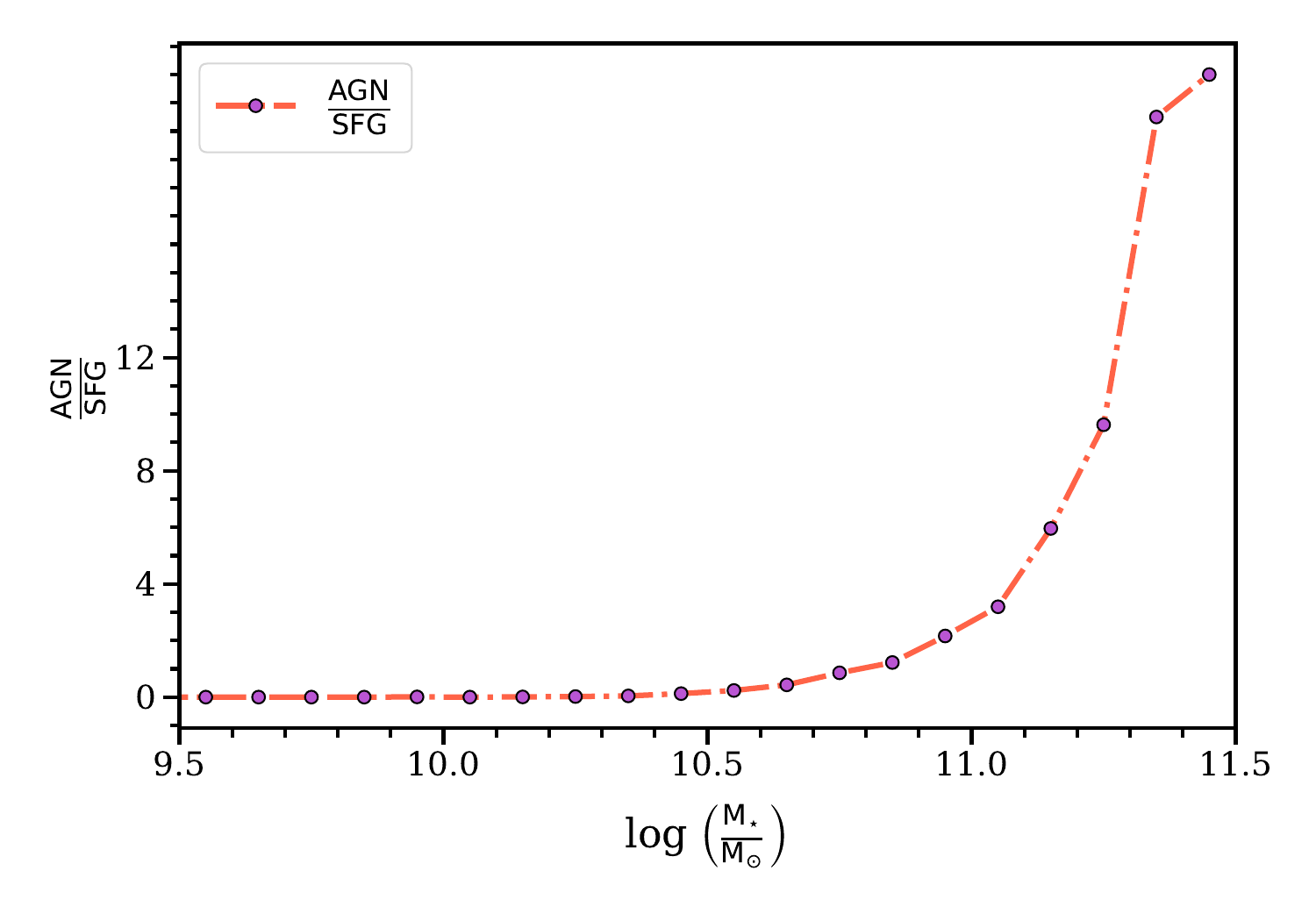}
    \caption{The left panel shows the fraction $\frac{AGN}{AGN+SFG}$
    and the right panel shows $\frac{AGN}{SFG}$ as a function of
    stellar mass.}
\label{fig:frac}
\end{figure*}

\begin{figure*}[htbp!]
    \centering
    \includegraphics[width = 8.5cm,height = 9.1cm]{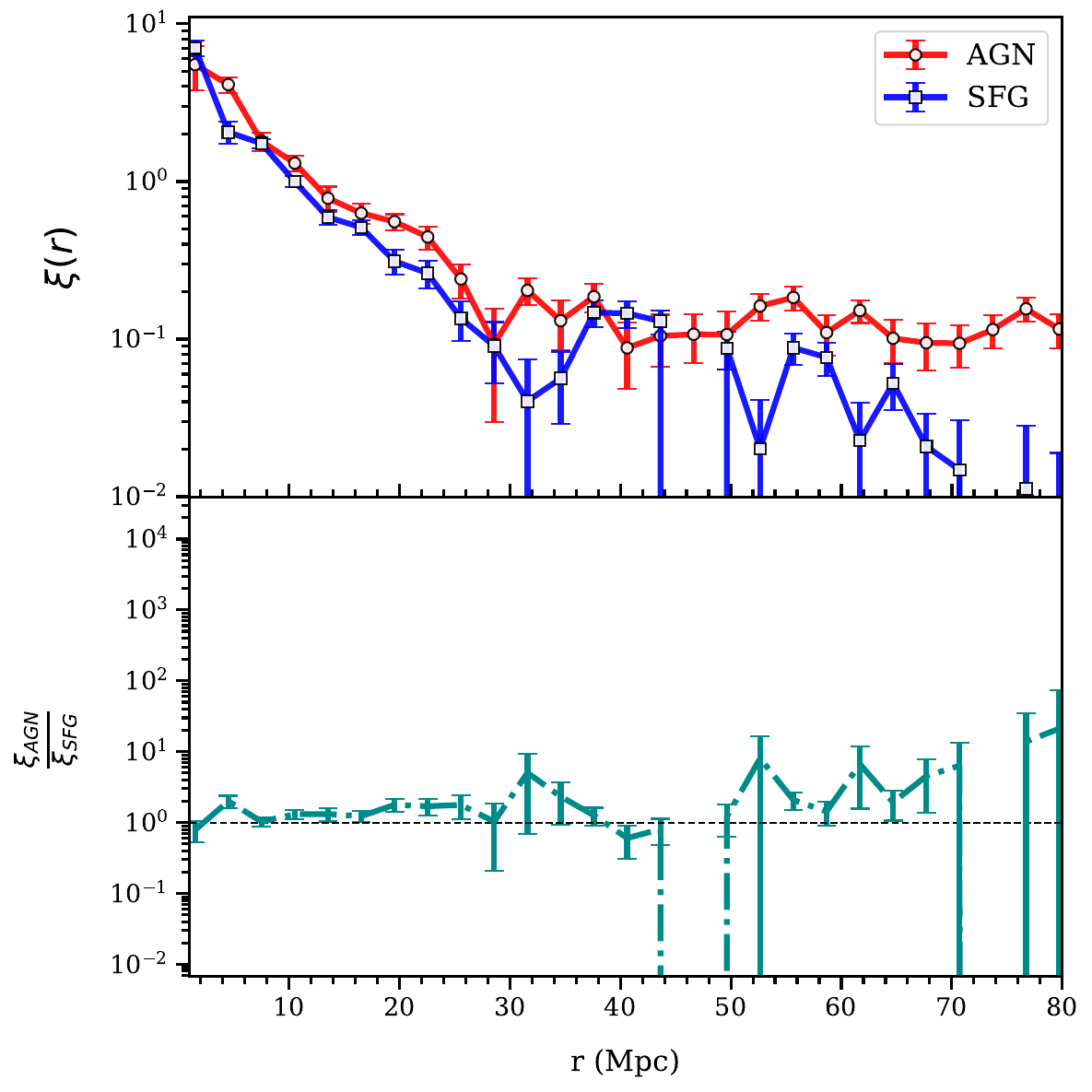}
    \vspace{-0.4cm}
    \includegraphics[width=9 cm,height=10.2cm]{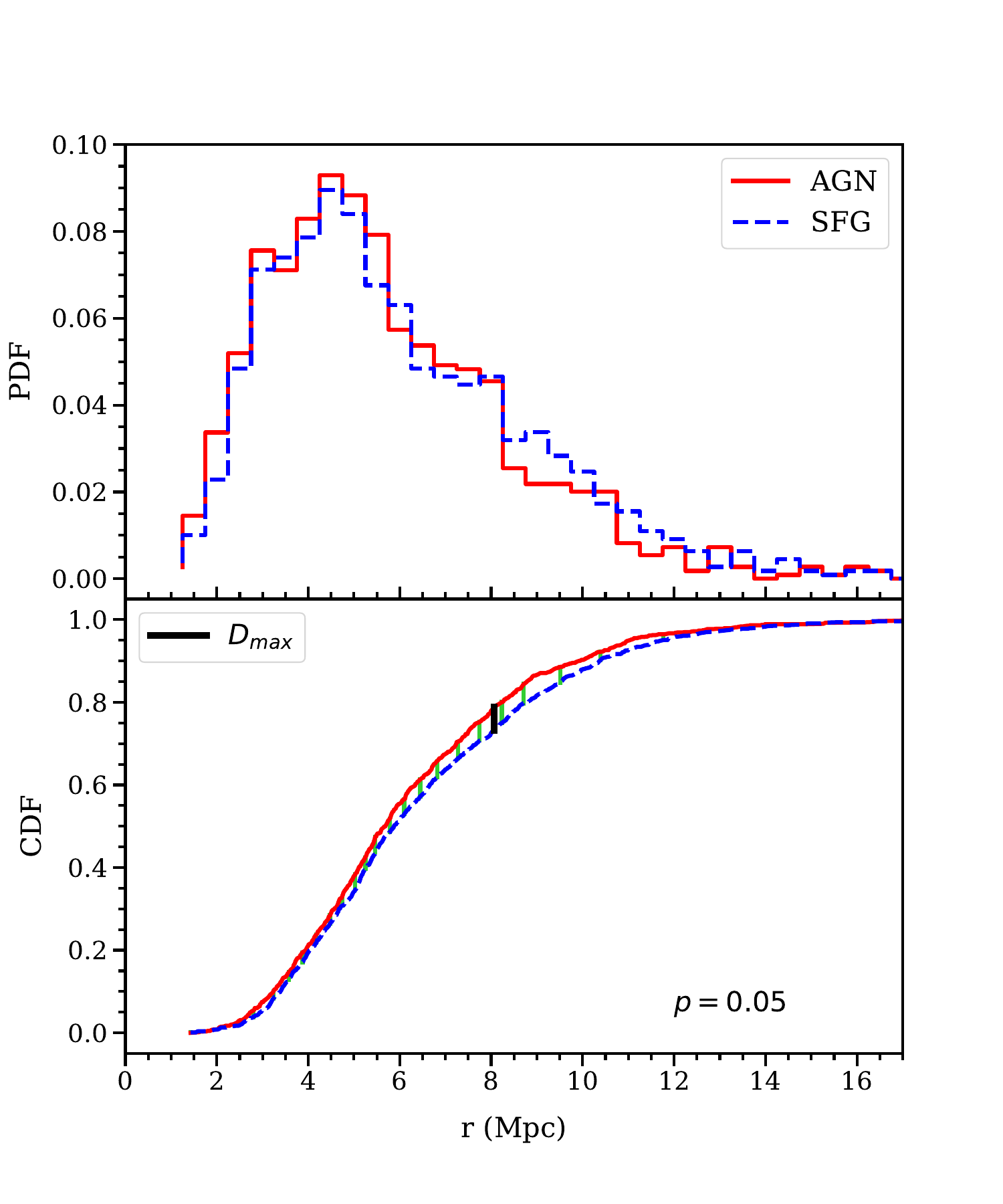}
    \caption{The top left panel of this figure shows the two-point
      correlation function as a function of length scale ($r$) for the
      AGN and SFG. The ratio of the two-point correlation functions
      for AGN and SFG is plotted as a function of $r$ in the bottom
      left panel. The 1$\sigma$ errorbars in these figures are
      obtained from $50$ jackknife samples drawn from the original
      dataset. The top right panel shows the PDFs of the $5^{th}$
      nearest neighbour distance for AGN and SFG. The two
      distributions are compared using a KS test, and the results are
      shown in the bottom right panel. The comparisons are carried out
      after matching the stellar mass distributions of AGN and SFG.}
\label{fig:clustmatch}
\end{figure*}
\hfill

% ****************************************************************** %

\renewcommand{\arraystretch}{1.5} % Increase row height by 1.5 times

\begin{table}
    \centering
    \begin{tabular}{|c|c|c|}
   \hline
        Class & Correlation length $(r_0)$ & Slope $(\gamma)$\\
        \hline
        AGN & $r_{0} = 10.82 \pm 3.41$ & $\gamma = 1.09 \pm 0.13$ \\
        \hline
        SFG & $r_{0} = 8.36 \pm 2.57$ & $\gamma = 1.29 \pm 0.12$ \\
        \hline
    \end{tabular}
   \caption{This table shows the best fit values of $r_{0}$ and
     $\gamma$ for the two-point correlation functions of AGN and
     SFG. The two-point correlation functions are fitted to a power
     law of the form $(\frac{r}{r_0})^{-\gamma}$ upto a scale of $~
     25$ Mpc.}
   \label{tab:fitparam}
\end{table}

% ****************************************************************** %
\subsection{The two-point correlation function and the $5^{th}$ nearest neighbour distribution of the mass-matched AGN and SFG}
\label{sec:2ptnn_pm}
In the top-left and top-right panels of \autoref{fig:clustmatch}, we
show the two-point correlation function and the probability density
function (PDF) of the $5^{th}$ nearest neighbour distance for the
mass-matched AGN and SFG populations, respectively. The bottom left
panel of \autoref{fig:clustmatch} reveals that AGN are somewhat more
strongly clustered than SFG at fixed stellar mass. However, the
statistical significance of these differences are not strong enough to
confirm the differences in their clustering strength. We also repeat
our calculations for the two-point correlation functions of AGN and
SFG using the publicly available code {\it Corrfunc} \citep{manodeep}
and obtained the same results as presented in this work.

The two-point correlation functions for AGN and SFG are analyzed in
redshift space, where a power-law fit provides a reasonable
approximation on scales below 25 Mpc \citep{hawkins03}.  We fit the
two-point correlation functions to a power law of the form
$\xi(r)=(\frac{r}{r_0})^{-\gamma}$ using least squares fitting and
present the fitted values for the correlation length ($r_0$) and slope
($\gamma$) in \autoref{tab:fitparam}. The results show that the
two-point correlation function of AGN has a larger correlation length
and a shallower slope compared to SFG, even after matching their
stellar mass distributions. However, the errors associated with these
parameters (see \autoref{tab:fitparam}) suggest that $r_0$ and
$\gamma$ for AGN are consistent with SFG within $1\sigma$.

The bottom-right panel of \autoref{fig:clustmatch} compares the
$5^{th}$ nearest neighbour distributions for AGN and SFG using a KS
test. The KS test shows that the null hypothesis can be rejected at
the $95\%$ confidence level. The distance to the $5^{th}$ nearest
neighbour is smaller for AGNs than for SFGs, indicating that AGN are
hosted in relatively higher-density regions compared to SFG. However,
the significance of these differences are not sufficiently strong that
can unambiguously provide an evidence in favour of a stronger
clustering of AGN compared to SFG. Several earlier studies reported a
stronger clustering for AGN \citep{gilli09, mandelbaum09, kollat12,
  donoso14, hale18}. Further investigations with larger datasets are
necessary to confirm the differences in the clustering of AGN host
galaxies and star-forming galaxies at fixed stellar mass.

% ****************************************************************** %
\begin{figure*}[htbp!]
    \centering
    \includegraphics[width = 7.5 cm]{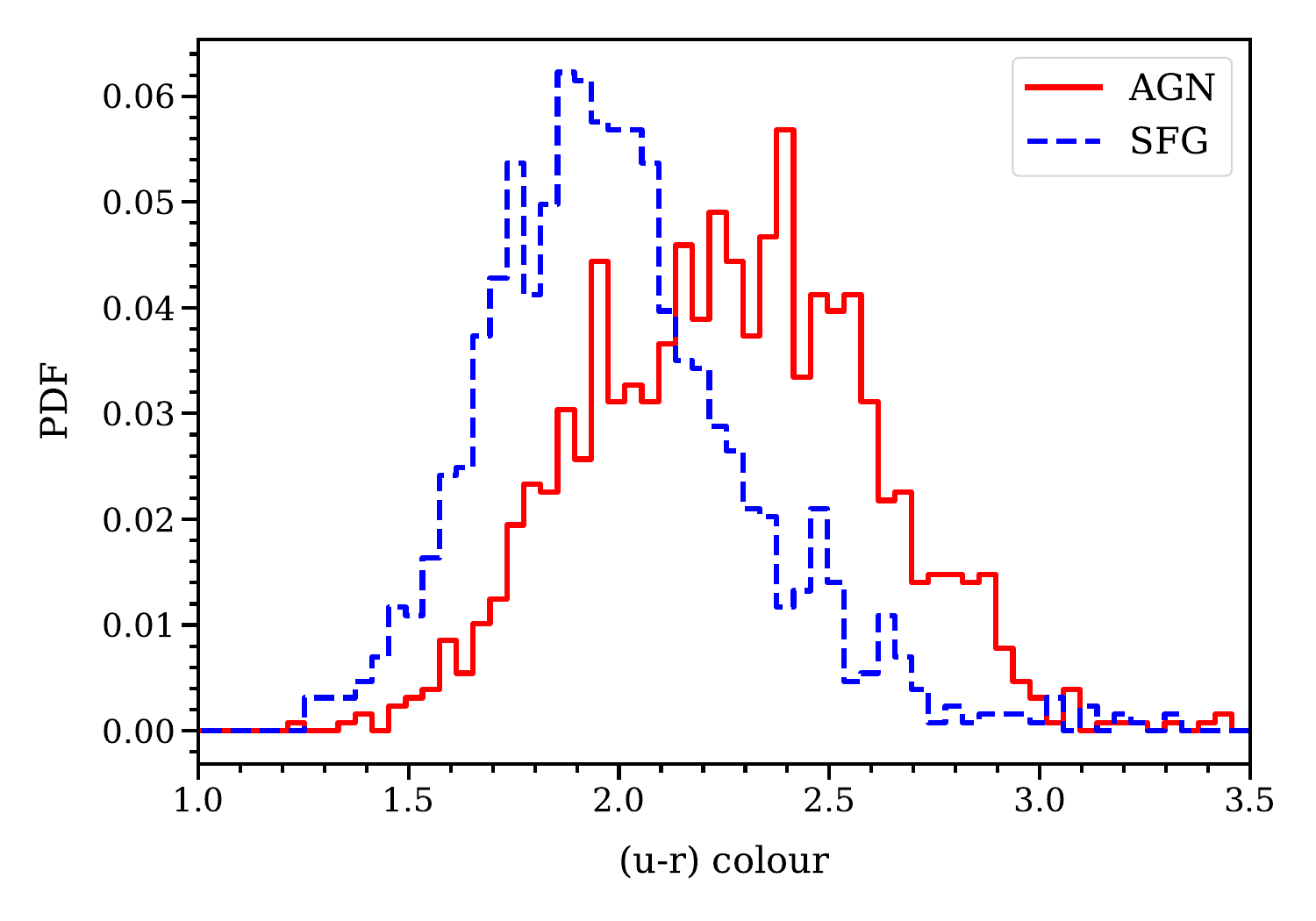}
    \vspace{-0.4cm}
    \includegraphics[width = 7.5 cm]{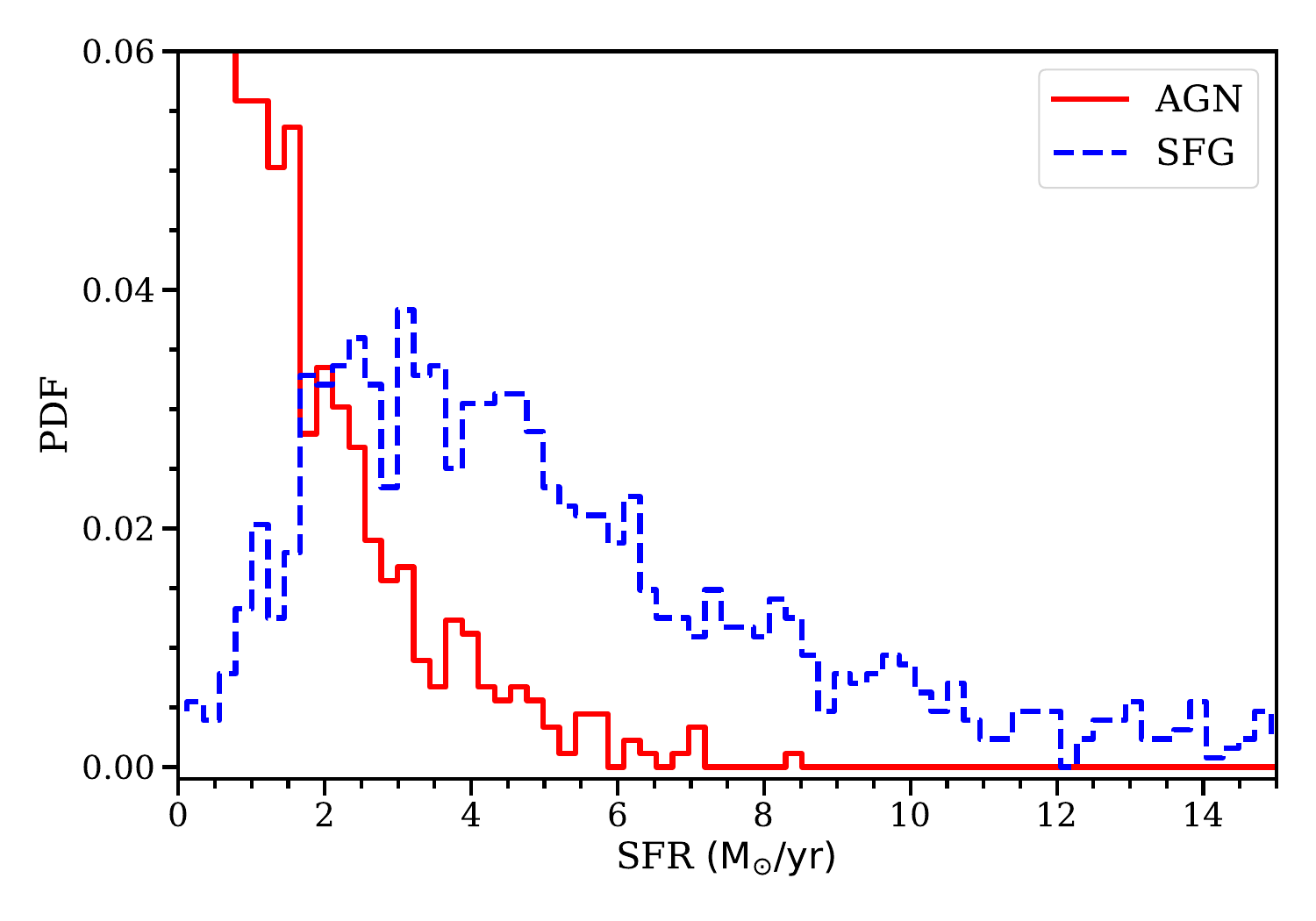}
    \vspace{-0.4cm}
    \includegraphics[width = 7.5 cm]{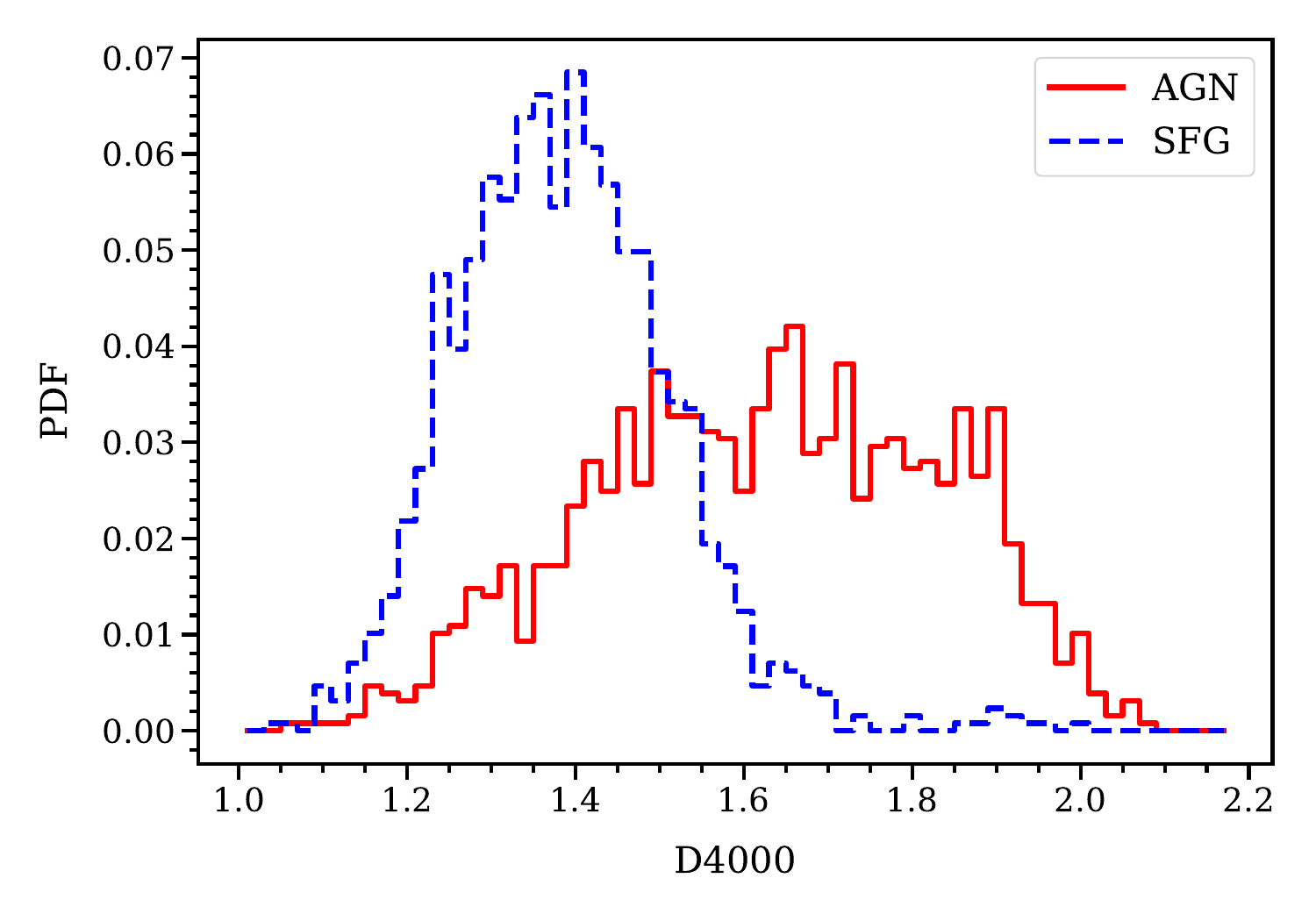}
    \vspace{-0.4cm}
    \includegraphics[width = 7.5 cm]{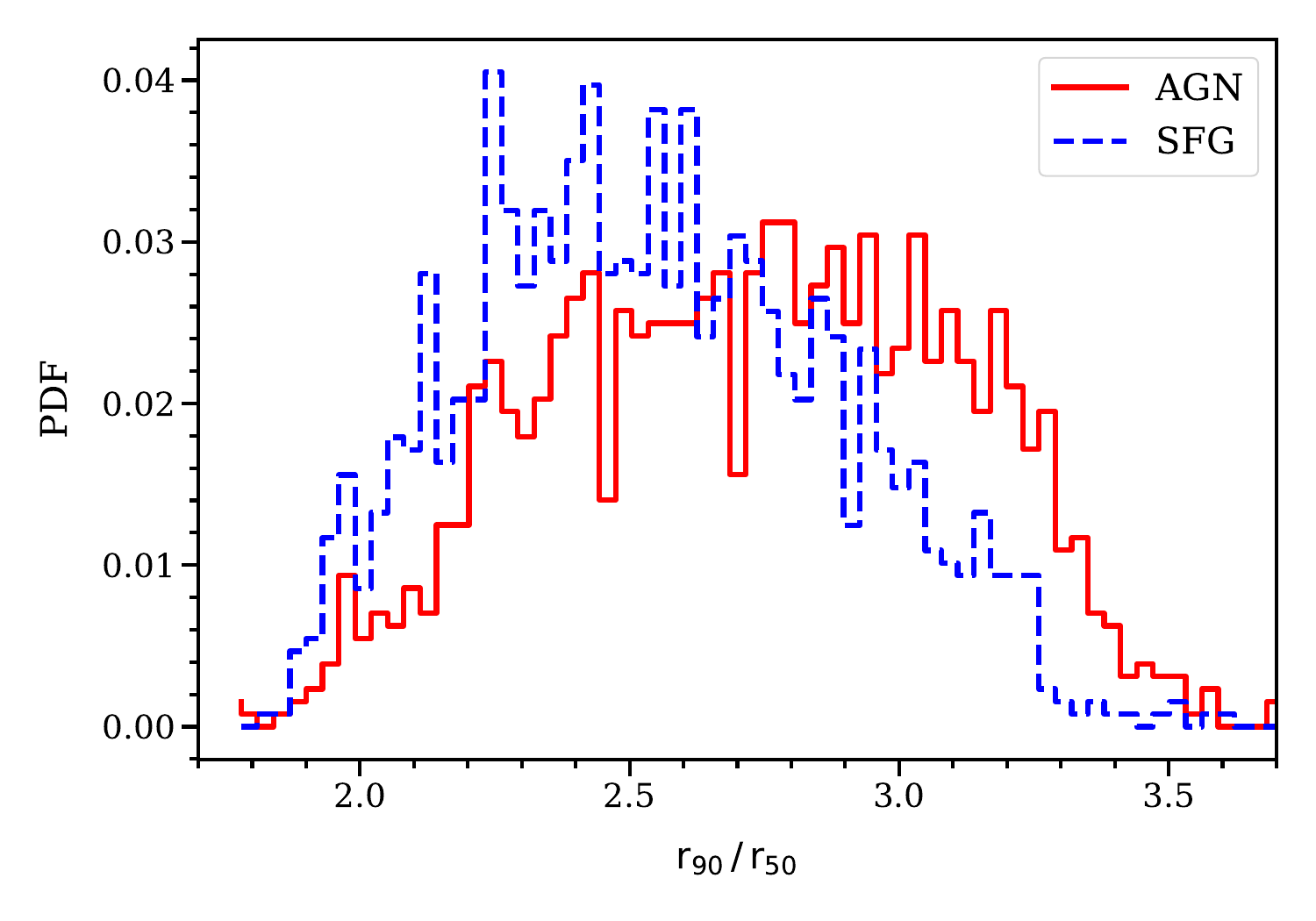}
    \vspace{0.5cm}
   \caption{The different panels of this figure show the distributions
     of the $(u-r)$ colour, SFR, D$4000$ and $\frac{r_{90}}{r_{50}}$
     for the AGN and SFG after matching their stellar mass
     distributions. We use KS-test to compare the distributions for
     the AGN and SFG in each case. The corresponding p-values are
     extremely small, and the null hypothesis can be rejected at
     $>99.99\%$ confidence level in each case.}
\label{fig:distmatch}
\end{figure*}

\begin{figure*}[htbp!]
\centering \includegraphics[width=7cm]{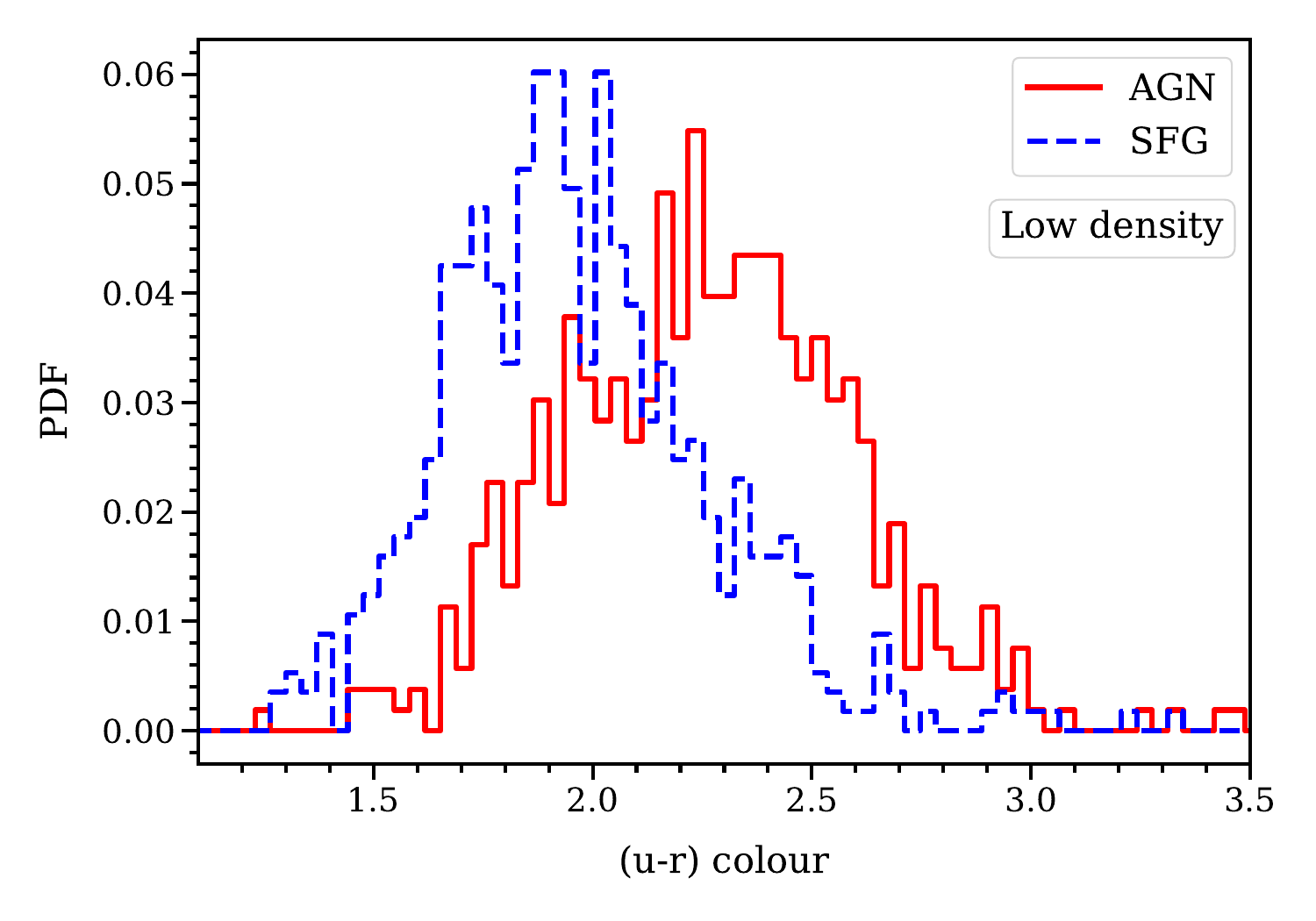}
\includegraphics[width=7cm]{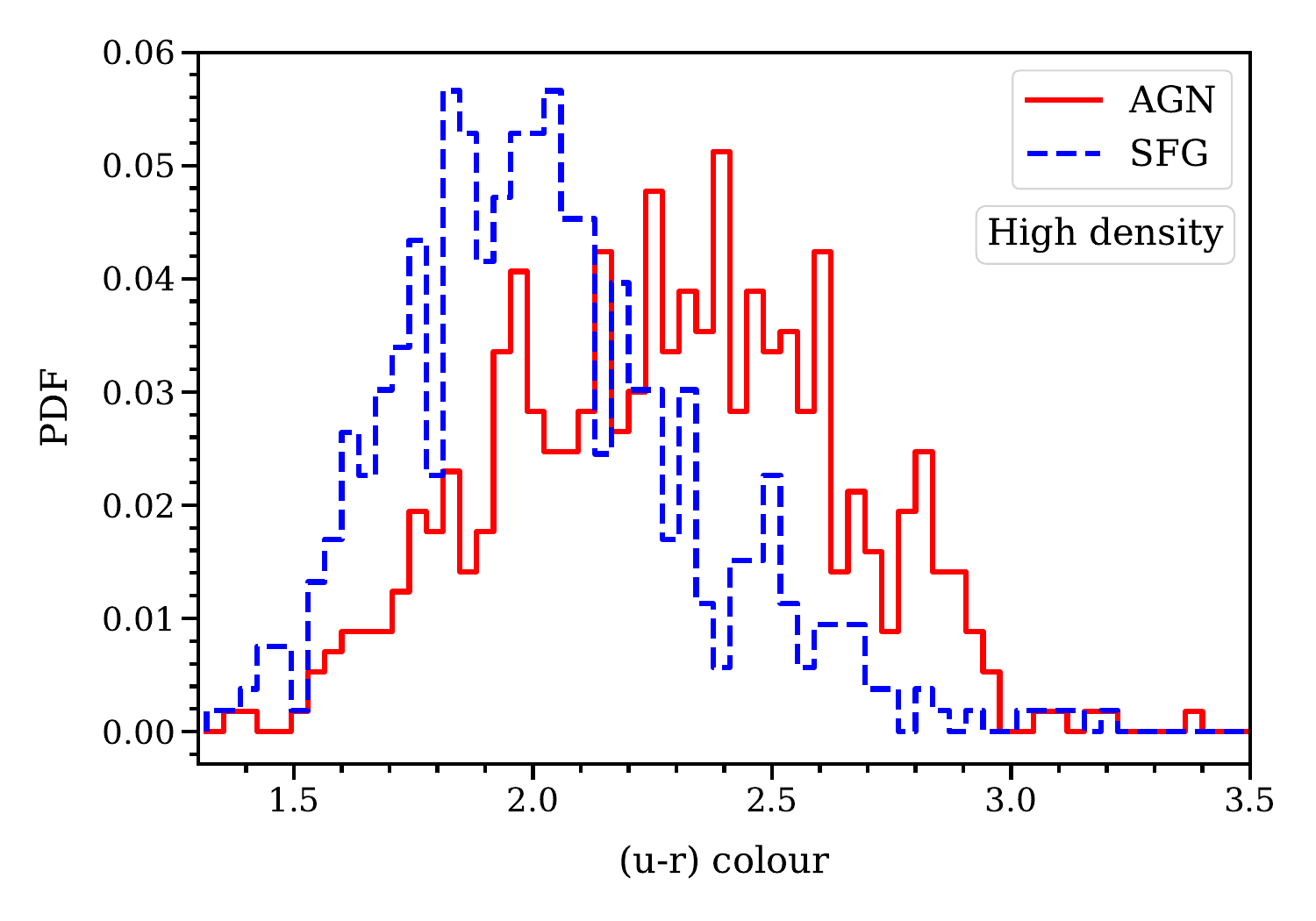}
\includegraphics[width=7cm]{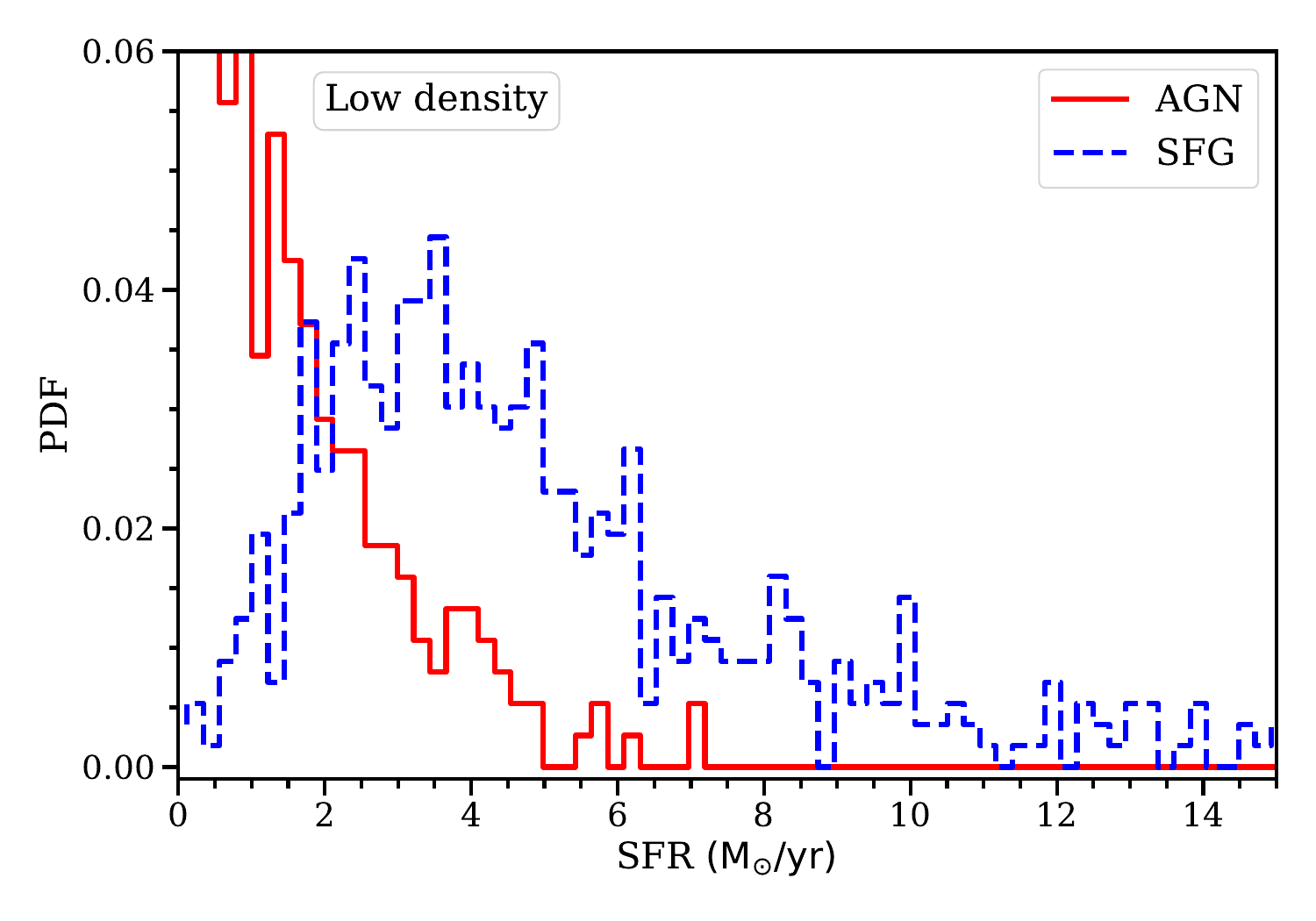}
\includegraphics[width=7cm]{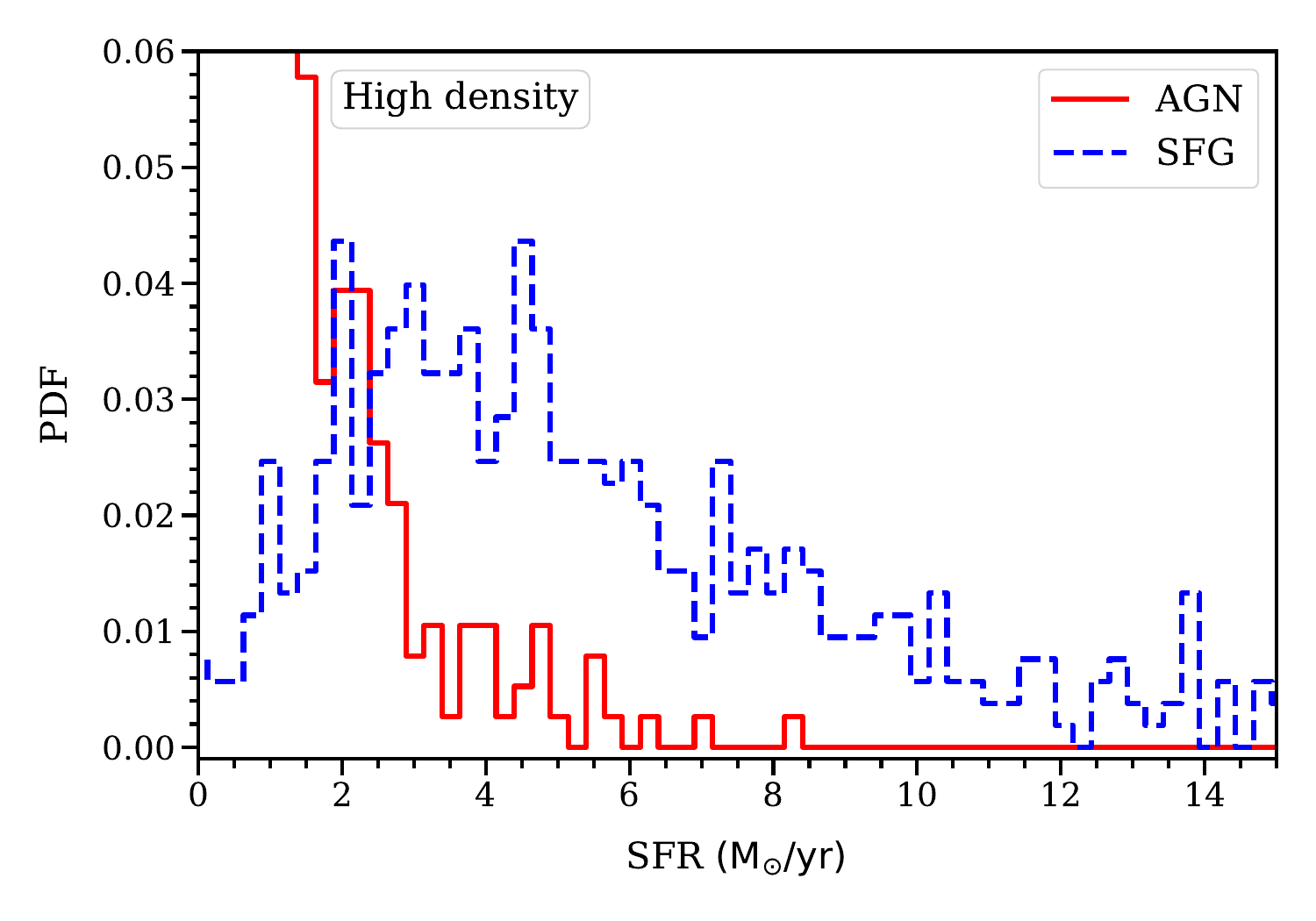}
\includegraphics[width=7cm]{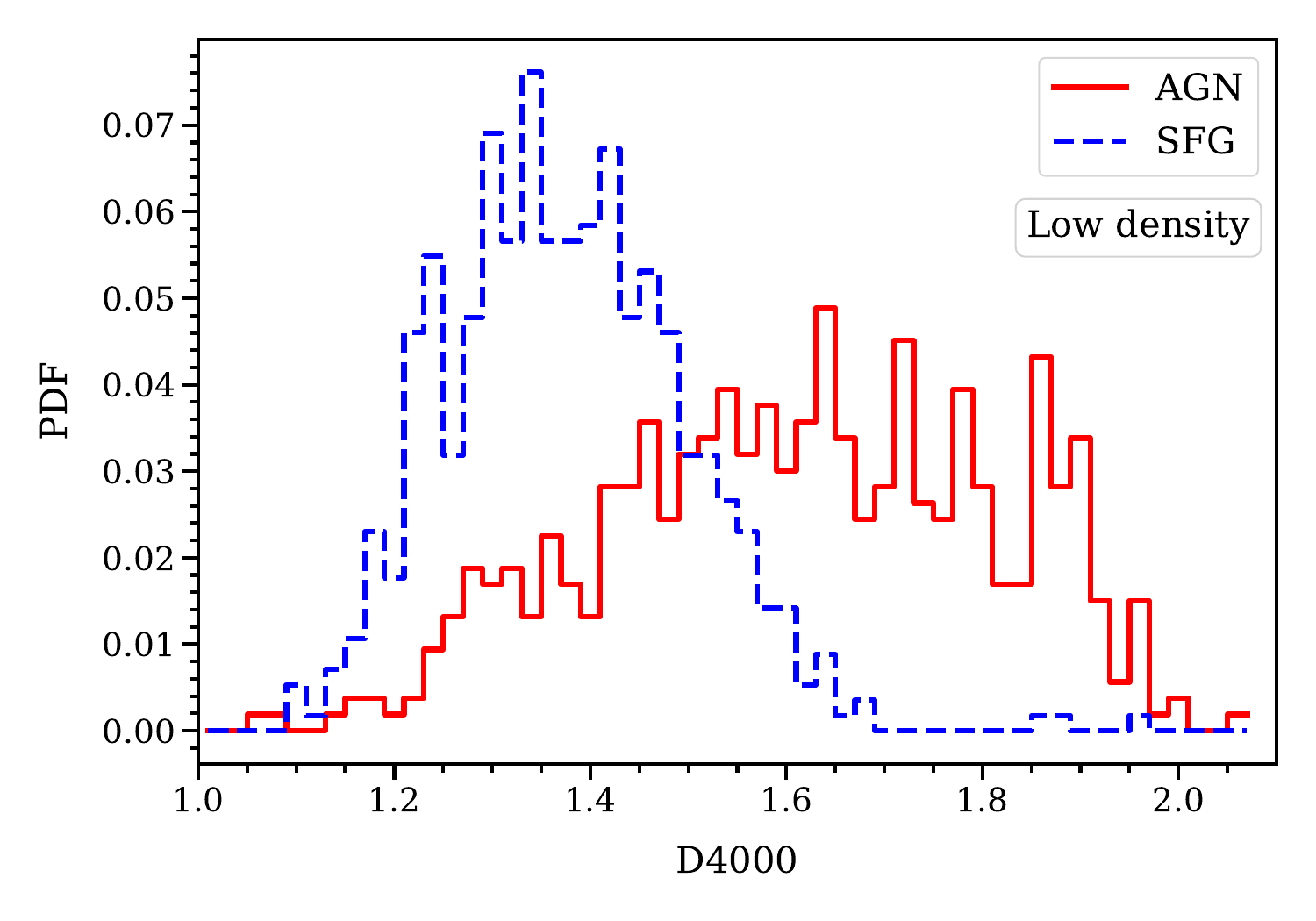}
\includegraphics[width=7cm]{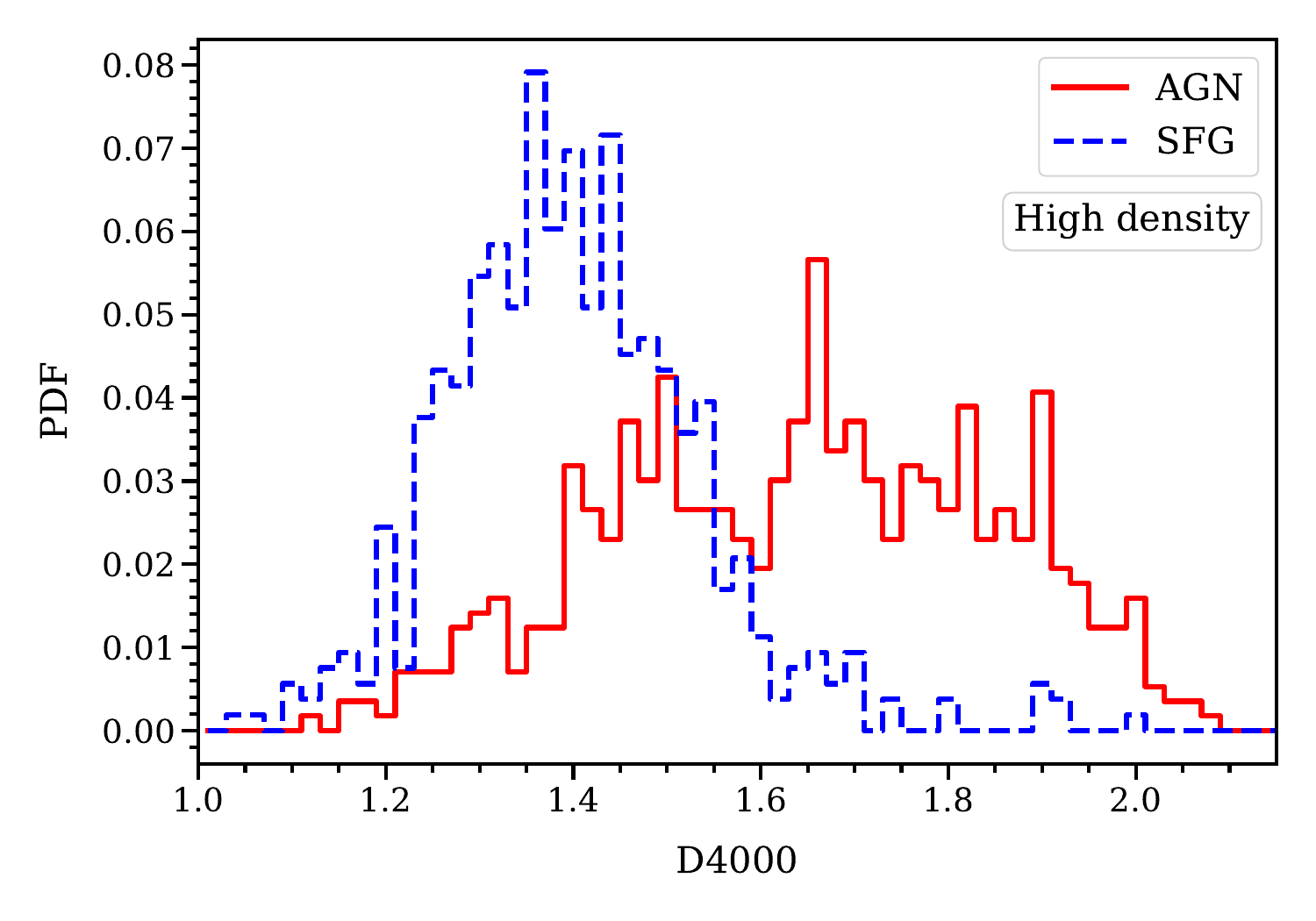}
\includegraphics[width=7cm]{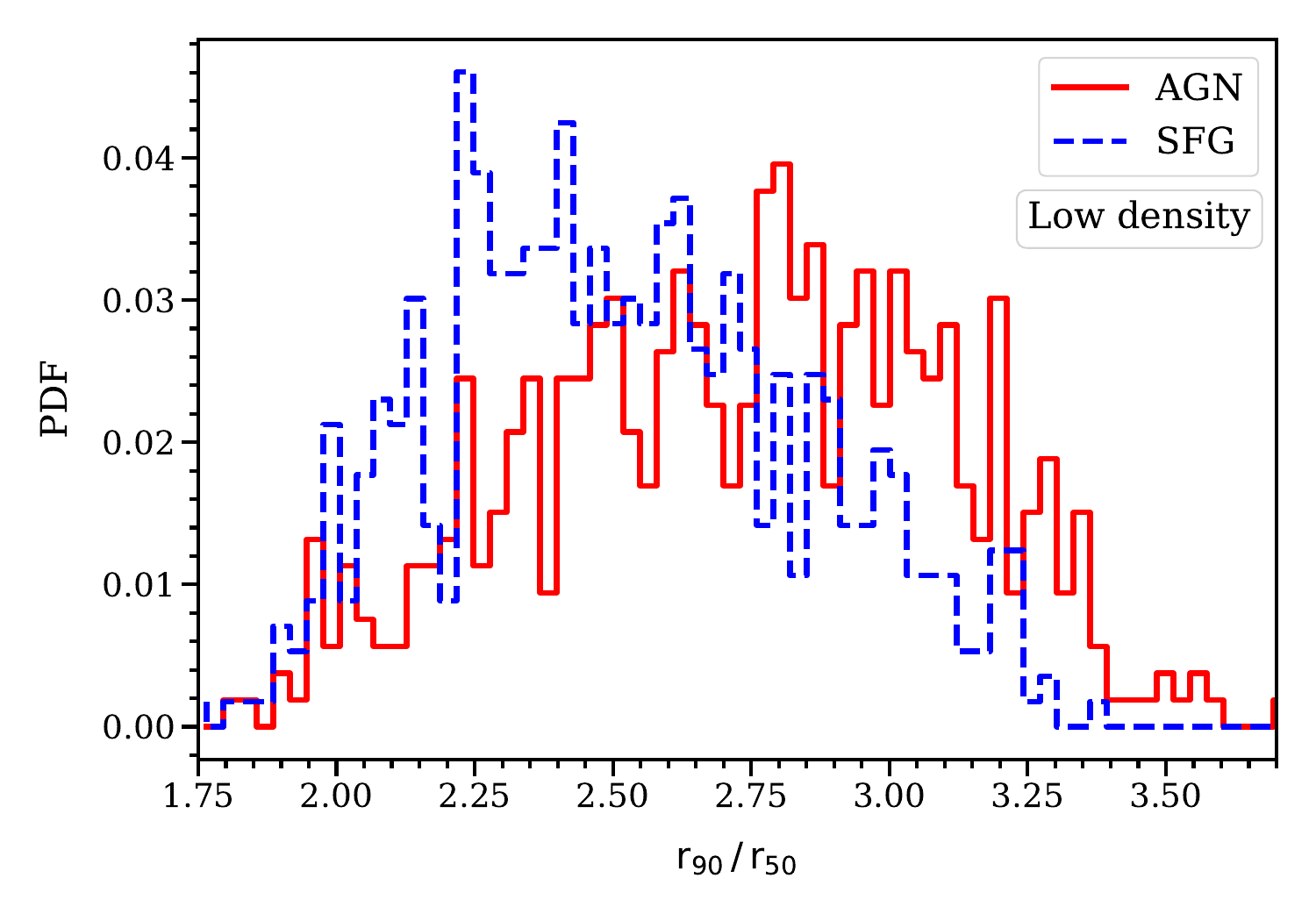}
\includegraphics[width=7cm]{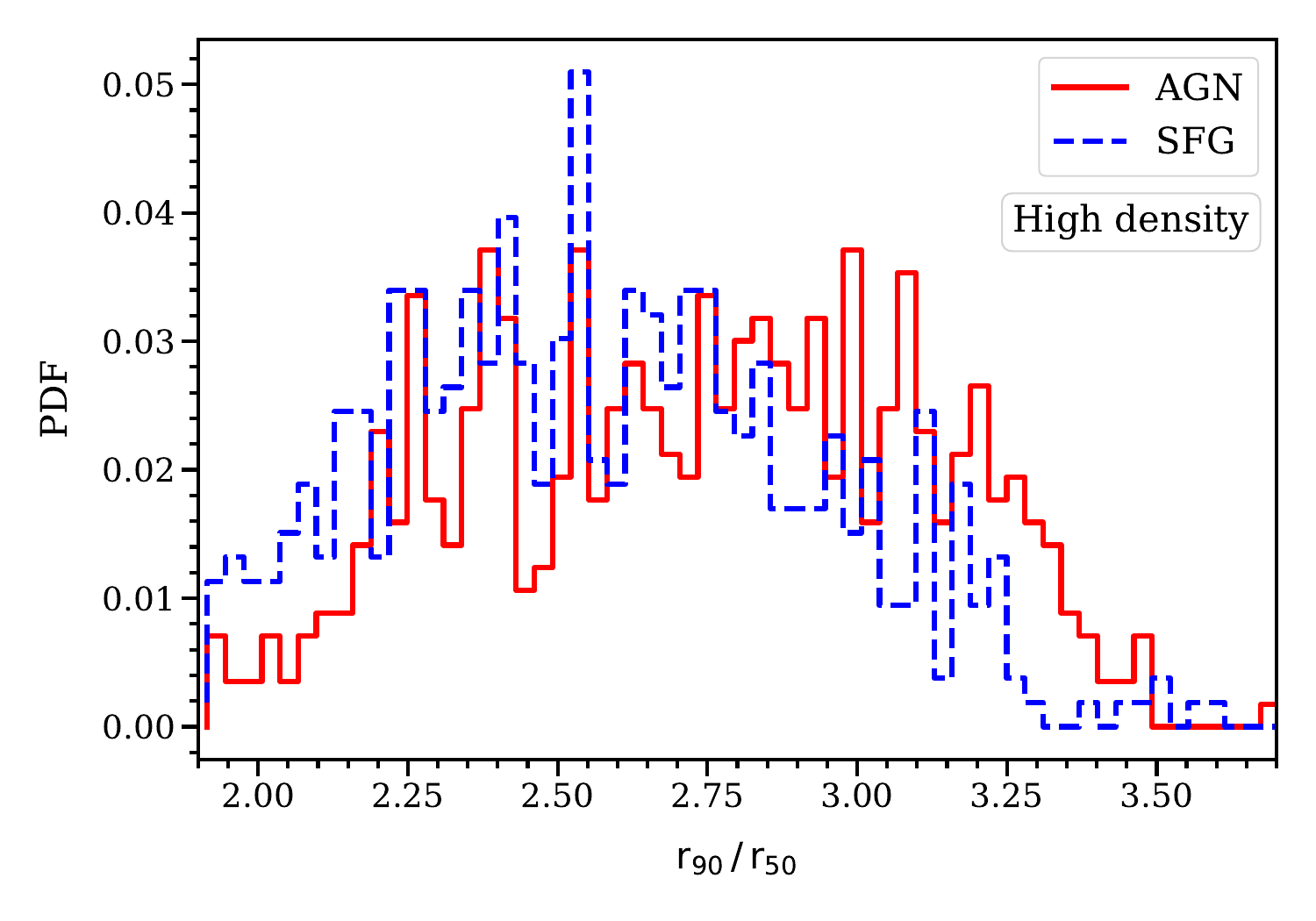}
\caption{The four left panels of this figure show the PDFs of $(u-r)$
  colour, SFR, D$4000$ and $\frac{r_{90}}{r_{50}}$ for the
  mass-matched AGN and SFG in the low density regions. The four right
  panels show the same in the high density regions. The KS test shows
  that the null hypothesis can be rejected at $>99.99\%$ confidence
  level in each case.}
\label{fig:denmatch}
\end{figure*}

% *********************************************************************************** %
\subsection{Comparing the distributions of different physical properties of the mass-matched AGN and SFG}
\label{sec:dist_pm}

The triggering of AGN activity may require specific physical
conditions within a galaxy, and the onset of AGN activity can, in
turn, affect certain physical properties of the host
galaxy. Understanding the differences between the physical properties
of AGN host galaxies and star-forming galaxies is crucial. The mass of
a galaxy is known to be the most influential factor in determining its
physical properties \citep{cooraysheth02}. Moreover, the AGN fraction
is strongly correlated with the stellar mass of galaxies
(\autoref{fig:frac}). It is therefore of interest to compare the
distributions of various physical properties for the two populations
after matching their stellar mass distributions.

We compare the distributions of the $(u-r)$ colour, concentration
index ($\frac{r_{90}}{r_{50}}$), SFR, and the D$4000$ for AGN host
galaxies and star-forming galaxies at fixed stellar mass. These
distributions are shown in different panels of
\autoref{fig:distmatch}. To quantify the dissimilarity between the two
distributions in each case, we apply the KS test. The results indicate
that the null hypothesis can be rejected with a confidence level
greater than $99.99\%$ in all cases.

The different panels of \autoref{fig:distmatch} show that the PDFs for
AGN and SFG cover similar ranges but peak at different values. In
the top-left panel of \autoref{fig:distmatch}, we observe that most
SFG are located in the blue cloud ($(u-r)<2.22$) \citep{strateva01},
while the colour distribution of AGN host galaxies peaks in the green
valley and extends into both the blue cloud and the red sequence.

The blue colours of SFG are primarily due to the presence of young,
hot, and massive stars that dominate the emission from the
galaxy. These stars emit substantial ultraviolet and blue light,
giving SFG their characteristic blue colour. In contrast, AGN tend to
have redder colours, which can be attributed to the dust and gas
surrounding the central black hole. This dust absorbs and scatters the
blue and ultraviolet light emitted by the accretion disk, causing the
galaxy to appear redder. Additionally, the redder colours may result
from the thermal emission of dust heated by the radiation,
contributing to the infrared part of the spectrum. An older stellar
population in AGN host galaxies can also contribute to their redder
appearance.

We compare the SFR distributions of AGN and SFG in the top-right
panel of \autoref{fig:distmatch}. The SFR distributions for SFG and
AGN peak around $\sim 4\, M_{\odot}/$yr and $\sim 1 \,M_{\odot}/$ yr,
respectively. Both distributions are positively skewed and extend to
higher SFRs (up to $15\, M_{\odot}/$ yr). However, the abundance of
AGN decreases significantly compared to SFG for SFRs above $3\,
M_{\odot}/$yr.

The bottom-left panel of \autoref{fig:distmatch} shows the
distributions of the $4000\textup{~\AA}$ break measurements for AGN
and SFG. The $4000\textup{~\AA}$ break is strongly correlated with
the ratio of the past average SFR to the present SFR in galaxies
\citep{kauffmann03a}, and it serves as an indicator of the galaxy's
recent star formation history. The distribution for AGN peaks at a
higher value of D$4000$ ($\sim 1.63$) and is negatively skewed, while
the distribution for SFG peaks around $\sim 1.38$ and is nearly
symmetrical. Lower values of D$4000$ ($<1.5$) are associated with
younger stellar populations, indicating recent star formation or a
completed starburst. Conversely, higher values ($>1.8$) correspond to
older stellar populations \citep{kauffmann09}. The higher D$4000$
values for AGN suggest that their host galaxies are primarily
composed of older stellar populations. However, we also observe that
some AGN host galaxies exhibit D$4000$ values below $1.5$, implying
that AGN activity can coexist with starburst activity in certain
galaxies.

We compare the distributions of the concentration index for AGN and
SFG in the bottom right panel of \autoref{fig:distmatch}. The
concentration index is strongly correlated with galaxy morphology
\citep{shimasaku01}. A concentration index of $\frac{r_{90}}{r_{50}} =
2.3$ corresponds to a pure exponential profile \citep{strateva01},
while $\frac{r_{90}}{r_{50}} = 3.33$ describes a pure de Vaucouleurs
profile \citep{blanton01}. Higher values of the concentration index
are typically associated with elliptical and bulge-dominated galaxies,
whereas disk-dominated spiral galaxies have lower concentration
indices ($< 2.6$) \citep{strateva01}. For our sample, the
concentration index distributions for AGN and SFG peak at around $\sim
2.8$ and $\sim 2.3$, respectively. This suggests that most SFG have
disk-like morphologies, while AGN are more commonly found in
bulge-dominated systems. We also note that the distribution for AGN is
negatively skewed, whereas the distribution for SFG is positively
skewed. This indicates that AGN can also occur in disk-dominated
galaxies, and some SFG may exhibit bulge-dominated morphologies. These
findings are consistent with previous studies showing that barred
spiral galaxies in groups often display AGN activity \citep{alonso14},
and that some elliptical galaxies can undergo rejuvenation in isolated
environments \citep{zezas03, lacerna16}.

% ********************************************************************************* %
\subsection{Comparing the distributions of different physical properties of the mass-matched AGN and SFG in low and high density regions}
\label{sec:den_pm}

\autoref{fig:distmatch} shows that the physical properties of AGN and
SFG differ significantly at fixed stellar masses. Analysis of the
two-point correlation function and the distribution of the $5^{th}$
nearest neighbours (\autoref{fig:clustmatch}) also reveals that AGN
exhibit moderately stronger clustering than SFG. AGN tend to prefer
denser regions, while SFG are more commonly found in less dense
environments. However, these environmental differences can not be
confirmed at a high significance level from this analysis. The local
density may have a role in triggering AGN activity. It would be
interesting to explore whether the observed differences in the
physical properties of AGN and SFG, as shown in
\autoref{fig:distmatch}, persist in regions of different density.  To
investigate this, we divide the mass-matched AGN and SFG into two
categories based on local density. Galaxies residing in regions with a
density below the median of the combined sample are classified as
``low density'', while those in regions with a density above the
median are classified as ``high density''.

We calculate the PDFs of four galaxy properties for AGN and SFG in
both low-density and high-density regions. The comparisons of physical
properties for AGN and SFG in low-density regions are shown in the
four left panels of \autoref{fig:denmatch}, while the comparisons in
high-density regions are displayed in the four right panels. The
differences between the PDFs in each panel are quantified using the
KS-test. The results show that the null hypothesis can be rejected
with a confidence level greater than $99.99\%$ in all cases,
indicating that the differences in the physical properties of AGN and
SFG persist in both low- and high-density regions.

AGN activity can be triggered in both high- and low-density
environments, and its presence significantly alters the physical
properties of the host galaxy compared to those of a SFG. Notably, the
differences in physical properties between AGN and SFG persist
regardless of local environmental density. This indicates that such
differences can not be explained by variations in local density.

% ================================================================================== %
\section{Conclusion}
We use a volume-limited sample from the SDSS to compare the clustering
and physical properties of SFG and AGN host galaxies at fixed stellar
mass. Our analysis with two-point correlation function and the
$5^{th}$ nearest neigbour distance reveals that the clustering
strength of AGN are moderately stronger than SFG. However, the
statistical significance of these differences are not sufficiently
strong to confirm these environmental differences. The weak
significance may arise due to the small size of our samples. Further
analysis with larger datasets are required for conclusive evidence.

We further compare the distributions of $(u-r)$ colour, concentration
index, SFR, and D$4000$ for AGN and SFG at fixed stellar mass and find
statistically significant differences at a confidence level exceeding
$99.99\%$. These distributions are also examined across varying
densities while maintaining fixed stellar mass, revealing that the
differences persist at the same significance level in both high and
low density environments (\autoref{fig:denmatch}). This suggests that
the observed differences in the physical properties of AGN and SFG
cannot be attributed solely to their local density. Instead, density
may play an indirect role in AGN activity by increasing the likelihood
of galaxy interactions \citep{ ellison11, sabater15, singh23}. In
relaxed systems, gas is unable to flow toward the central SMBH due to
angular momentum conservation. Interactions can generate torques or
instabilities that funnel gas toward the SMBH, thereby triggering AGN
activity \citep{woods07, rogers09}. Although the number density of
galaxies in cluster environments is much higher than in the field, the
higher velocities of galaxies, particularly those newly infalling near
the cluster center, can inhibit interactions.  \citet{haines12}
provide evidence suggesting that galaxy interactions may still play an
important role in the cluster outskirts, offering a scenario where
such interactions are more likely to occur. \citet{ehlert15} suggested
that galaxy mergers could play a significant role in contributing to
the AGN population within clusters. Several other works
\citet{koulouridis24, drigga25} presented additional evidence in
support of this idea. Notably, our results indicate that AGN activity
can also be sustained through secular processes in low density
environments.

Galaxies with similar stellar masses can exhibit significantly
different assembly histories, creating uncertainties about whether a
galaxy of a given stellar mass can host the conditions necessary for
AGN activity. Two key prerequisites for AGN activity are the presence
of a bulge and the availability of gas \citep{ruffa19, shangguan20,
  ellison21, sampaio23}. However, these favorable conditions are met
only in a subset of galaxies at a given stellar mass, with their
prevalence depending on both stellar mass and assembly
history. Studies indicate that the frequency of bulge formation
increases with stellar mass \citep{erwin17} and is influenced by
assembly history \citep{kruk19}. Hydrodynamical simulations further
suggest that assembly bias can lead to substantial variations in the
cold gas content of galaxies \citep{cui21}. For instance, galaxies
with higher stellar masses tend to reside in early-formed halos, which
are more likely to accumulate large reservoirs of cold
gas. Consequently, the availability of cold gas is governed by both
stellar mass and the assembly history of the host halos.

The observed increase in AGN fraction with stellar mass is linked to
the greater likelihood of bulge dominance and the presence of larger
cold gas reservoirs in more massive galaxies. Additionally, massive
halos, which reside in denser environments, are subject to more
frequent interactions. These interactions can influence AGN activity
and may be reflected in the clustering properties of
galaxies. \cite{croton07} demonstrate that assembly bias alters the
two-point correlation function of galaxies by $\sim 10\%$. It would be
difficult to confirm the role of assembly bias from such small
differences in the observed two-point correlation functions. Using
information theoretic measures, several studies show that galaxy
morphology and colour are significantly impacted by large-scale
environment \citep{pandey17, sarkar20, sarkar22}. Such dependence
hints towards possible roles of assembly bias in shaping the galaxy
properties. We plan to compare the large-scale environment of SFG and
AGN in future work.

Massive galaxies predominantly inhabit denser environments, and AGN
hosts are typically high-mass galaxies. Since galaxy clustering is
strongly influenced by mass, our SFG and AGN samples are matched in
stellar mass. AGN exhibit only a moderately stronger clustering than
SFG, suggesting that local density may not have a significant role in
triggering AGN activity. Further, the observed differences in the
physical properties of mass-matched SFG and AGN remain largely
independent of environmental density. This also indicates that the
local environment is unlikely to be the primary driver of AGN
activity.

Our findings hint at a potential role of assembly history in
influencing AGN activity. However, the current analysis does not
provide conclusive evidence, necessitating further investigation. The
relationship between assembly bias and AGN activity is inherently
complex, involving an intricate interplay between the formation
histories of galaxies, the properties of their host dark matter halos,
and the mechanisms that trigger and regulate AGN activity. A deeper
understanding of these connections is essential to uncover the
influence of assembly bias on AGN activity. We plan to use
hydrodynamical simulations, such as EAGLE \citep{schaye15} and
IllustrisTNG \citep{nelson19} in future work, to explore these aspects
in greater detail.

\begin{acknowledgement}
The authors thank an anonymous reviewer and the associate editor for the valuable comments and
suggestions that helped to improve the draft.  BP would like to
acknowledge financial support from the SERB, DST, Government of India
through the project CRG/2019/001110.  BP would also like to
acknowledge IUCAA, Pune, for providing support through the
associateship programme. AN acknowledges the financial support from
the Department of Science and Technology (DST), Government of India
through an INSPIRE fellowship. The authors thank Tapas Kumar Das for
some interesting discussions.

Funding for the SDSS and SDSS-II has been provided by the Alfred
P. Sloan Foundation, the Participating Institutions, the National
Science Foundation, the U.S. Department of Energy, the National
Aeronautics and Space Administration, the Japanese Monbukagakusho, the
Max Planck Society, and the Higher Education Funding Council for
England. The SDSS website is \url{http://www.sdss.org/}.

The SDSS is managed by the Astrophysical Research Consortium for the
Participating Institutions. The Participating Institutions are the
American Museum of Natural History, Astrophysical Institute Potsdam,
University of Basel, University of Cambridge, Case Western Reserve
University, University of Chicago, Drexel University, Fermilab, the
Institute for Advanced Study, the Japan Participation Group, Johns
Hopkins University, the Joint Institute for Nuclear Astrophysics, the
Kavli Institute for Particle Astrophysics and Cosmology, the Korean
Scientist Group, the Chinese Academy of Sciences (LAMOST), Los Alamos
National Laboratory, the Max-Planck-Institute for Astronomy (MPIA),
the Max-Planck-Institute for Astrophysics (MPA), New Mexico State
University, Ohio State University, University of Pittsburgh,
University of Portsmouth, Princeton University, the United States
Naval Observatory, and the University of Washington.
\end{acknowledgement}

\paragraph{Data Availability Statement}
The SDSS data are publicly available at
https://skyserver.sdss.org/casjobs/. The data generated in this work
will be shared on reasonable request to the authors.

\printbibliography

\end{document}